\documentclass[12pt]{article}

\usepackage{epsfig}

\usepackage{amssymb}
\usepackage{amsfonts}

\usepackage{color}
 
\def\be{\begin{equation}}
\def\ee{\end{equation}}
\def\ba{\begin{array}{c}}
\def\ea{\end{array}}

\def\ben{$$}
\def\een{$$}

\newcommand{\bea}{\begin{eqnarray}}
\newcommand{\eea}{\end{eqnarray}}

\newcommand{\pkt}{\!\!\succ\,\,}
\newcommand{\kt}{\rangle}
\newcommand{\br}{\langle}

\newtheorem{thm}{Theorem}

\newtheorem{lemma}[thm]{Lemma}

\begin{document}

\begin{center}

.

{\Large \bf

Systematics of quasi-Hermitian representations
of non-Hermitian quantum models

}

\vspace{10mm}

\vspace{0.2cm}

\vspace{2mm}

\textbf{Miloslav Znojil}

\vspace{0.2cm}

\vspace{0.2cm}

Department of Physics, Faculty of Science, University of Hradec
Kr\'{a}lov\'{e},

Rokitansk\'{e}ho 62, 50003 Hradec Kr\'{a}lov\'{e},
 Czech Republic

\vspace{0.2cm}

 and

\vspace{0.2cm}

The Czech Academy of Sciences, Nuclear Physics Institute,

 Hlavn\'{\i} 130,
250 68 \v{R}e\v{z}, Czech Republic

\vspace{0.2cm}

{e-mail: znojil@ujf.cas.cz}

\end{center}

\newpage

\section*{Abstract}

In the currently quickly growing area of applications of
non-Hermitian Hamiltonians $H\neq H^\dagger$ which are quasi-Hermitian
(i.e., such that $H^\dagger\, \Theta=\Theta\,H$, with a suitable
inner-product metric $\Theta\neq I$),
the correct probabilistic interpretation of the model
is needed.
Achieved either in the
Buslaev-Grecchi-inspired
spirit (BGI;
one factorizes $\Theta=\Omega^\dagger\,\Omega$ and reconstructs
the conventional
Hermitian Hamiltonian,
$H \to \mathfrak{h} = \Omega\,H\,\Omega^{-1}=\mathfrak{h}^\dagger$)
or in the much user-friendlier,
Dyson-inspired spirit (DI; one
eliminates
the ``difficult''
reference to $\mathfrak{h}$ via the
quasi-Hermiticity rule, i.e.,
via an ``easier'' reconstruction of $\Theta$).
Here, the two model-building BGI and DI recipes
are identified as the two extreme special cases
of a general correct-interpretation-providing
strategy. We show that at any preselected integer $N$
one may make a choice
between
the BGI extreme
and an $N-$plet of the other, consistent and physical
DI-type
representations of the system in which
a specific, partially modified Hamiltonian
is constructed as
quasi-Hermitian with respect to a specific,
simplified inner-product metric. In applications,
any one of these $N+1$ options
may prove optimal for a given $H$:
A schematic three-state quantum system
is discussed as an illustrative example.


\section*{Keywords}
.

non-Hermitian quantum mechanics of unitary systems;

isospectral preconditionings of Hamiltonians;

physical Hilbert spaces with ad hoc inner-product metrics;

alternative reconstructions of
hidden Hermiticity;

\newpage

\section{Introduction}

In the conventional
formulation of quantum mechanics
in Schr\"{o}dinger picture \cite{Messiah}
one has to solve
Schr\"{o}dinger equation
 \be
 {\rm i}\,\frac{d}{dt}\, |\psi\pkt
 = \mathfrak{h}\, |\psi\pkt\,,
 \ \ \ \ \
 \mathfrak{h}=\mathfrak{h}^\dagger\,,
 \ \ \ \ \ |\psi\pkt \in {\cal H}_{textbook}\,.
 \label{stasta}
 \ee
This process
is often facilitated by a unitary isospectral preconditioning
of the Hamiltonian,
 \ben
 \mathfrak{h}\, \to \, \mathfrak{h}'= {\cal U}^\dagger\, \mathfrak{h}\, {\cal U}
 \,,\ \ \ \ \
 \mathfrak{h}'=\left (\mathfrak{h}'\right )^\dagger
 \,.
 \een
In 1956
F. Dyson revealed,
during his study of ferromagnetism \cite{Dyson},
that the efficiency of the preconditioning
can decisively be enhanced when one omits the
formally redundant condition of the unitarity of the transformation.
Then, one generalizes
 \be
 \mathfrak{h} \to  H= \Omega^{-1}\, \mathfrak{h}\, \Omega\,,
 \ \ \ \ \ \Omega^\dagger\,\Omega \neq I
 \,,\ \ \ \ H \neq
  \,H^\dagger\,
 \label{preco}
 \ee
and
speaks about the Dyson-inspired (DI) ``non-Hermitian Hamiltonian''
reformulation of the conventional quantum mechanics
(for its compact outline see Appendix A below).

In the latter framework
the original
Schr\"{o}dinger Eq.~(\ref{stasta}) for
$ |\psi\pkt=\Omega\,|\psi\kt$
is merely replaced
by its mathematically equivalent non-Hermitian alternative
 \be
 {\rm i}\,\frac{d}{dt}\, |\psi\kt
 = H\, |\psi\kt\,,
 \ \ \ \ \ |\psi\kt \in {\cal H}_{math}\,
 \label{ostasta}
 \ee
which is, presumably,
easier to solve.
Naturally, the latter assumption is essential and nontrivial because
the amended representation $H$ of the
Hamiltonian becomes less standard, viz.,
quasi-Hermitian \cite{Dieudonne,Geyer},
 \be
 H^\dagger\,\Theta = \Theta\,H\,,
 \ \ \ \ \ \Theta=\Omega^\dagger\Omega\,.
 \label{starstar}
 \ee
In practice, nevertheless,
the generalized DI preconditioning (\ref{preco})
{\it alias\,} ``Dyson map'' {\it alias\,} ``quasi-Hermitian
quantum mechanics'' (QHQM)
found several successful
applications, say, in nuclear physics
(see, e.g., \cite{Jenssen,Navr} and also \cite{Bishop}).

On these grounds, Scholtz et al  \cite{Geyer} proposed, in 1992,
a closely related modification of the conventional
model-building process.
The essence of their innovative proposal lied in a reversal
of the
preconditioning,
 \be
 H \to \mathfrak{h} = \Omega\, H\, \Omega^{-1}= \mathfrak{h}^\dagger\,.
  \label{repreco}
 \ee
In place of the traditional approach
in which one recalls the principle of
correspondence and
in which one preselects a self-adjoint candidate $\mathfrak{h}$
for the Hamiltonian, an alternative, textbooks-complementing
upgrade of QHQM in
Schr\"{o}dinger picture has been
described and promoted.
In this approach
the non-Hermitian but still user-friendly ``input''
Hamiltonian $H$
is expected to be given in advance.

\begin{table}[h]

\caption{The DI-Hermitization pattern
${\cal H}_{math}\to{\cal H}_{phys}$.}
 \label{os1x} \vspace{.4cm}
\centering
\begin{tabular}{||c|c|c|c|c||}
    \hline \hline
    input:
& aim:& construction: & tool:
&result: \\ Hamiltonian
 & unitarity & find the metric& amend the products
& Hermiticity  in ${\cal H}_{phys}$
 \\ $H \neq H^\dagger$ in ${\cal H}_{math}$
 & (in ${\cal H}_{phys}$) & $\Theta=\Theta(H)$
&$\br \psi|\psi'\kt\to\br \psi|\Theta|\psi'\kt$
 &$H=\Theta^{-1}\,H^\dagger\,\Theta \equiv H^\sharp$
 \\
\hline \hline
\end{tabular}
\end{table}

Naturally,
the preselected operator $H$ must be
assigned the obligatory probabilistic
physical interpretation \cite{MZbook}.
The purpose can be served either by
a direct backward reconstruction
of $\mathfrak{h}$
(via Eq.~(\ref{repreco}))
or
by the direct reference to the
quasi-Hermiticity of $H$.
In the latter, DI setting
(see its compact outline
in Table \ref{os1x})
the construction is
usually more straightforward because
one merely has to solve
Eq.~(\ref{starstar}) in order to obtain
the so called metric operator $\Theta=\Theta(H)$
(the technical details
may be found outlined, e.g., in reviews \cite{Geyer,ali}).
Another reason of the preference of the DI strategy in applications
lies in its conceptual simplicity
because the initial non-Hermitian
Hamiltonian $H$ may be kept unchanged.
In many applications
it is not even necessary to know or reconstruct
the Dyson map.

For a sample
of applicability of the former,
Hamiltonian-amending approach
one may recall
the 
realistic anharmonic-oscillator
example as described, in 1993,
by Buslaev and Grecchi \cite{BG}.
This result
(see its account in Appendix~B below)
constituted, in fact, also a key source of inspiration
of our present study.
Nevertheless,
one has to admit
that
in comparison with the DI quasi-Hermitization,
the straightforward Buslaev-Grecchi-inspired (BGI)
Hermitization (\ref{repreco})
(where
one must obligatorily re-factorize
the metric $\Theta=\Omega^\dagger\Omega$, see Table \ref{os2x})
has to be perceived as technically less friendly.

\begin{table}[h]
\caption{BGI Hermitization
${\cal H}_{math}\to{\cal H}_{textbook}$.}
 \label{os2x} \vspace{.4cm}
\centering
\begin{tabular}{||c|c|c|c|c||}
    \hline \hline
    input:
& aim:& prepartory step: & main step:
&result: \\ Hamiltonian
 & unitarity & find & decompose
& Hermiticity in ${\cal H}_{textbook}$
 \\ $H \neq H^\dagger$ in ${\cal H}_{math}$
 & (in ${\cal H}_{textbook}$) & $\Theta=\Theta(H)$
&$\Theta=\Omega^\dagger\Omega$
 &$\mathfrak{h}=\Omega\,H\,\Omega^{-1}=\mathfrak{h}^\dagger$
 \\
\hline \hline
\end{tabular}
\end{table}

For compensation,
it makes sense to point out that
the Buslaev's and Grecchi's constructive
Hermitization (\ref{repreco})
was rendered possible by a hidden symmetry of their
non-Hermitian model
$H^{(BG)}$ with respect to a family of
partial Dyson sub-maps $\Omega_j$.
This opens the way towards an innovative
$N-$term factorization
of the global Dyson-map operator itself,
 \be
 \Omega=\Omega_N\Omega_{N-1}\ldots\Omega_1\,.
 \label{cocody}
 \ee
One of the main goals of our present analysis
will be, therefore, a model-independent, abstract clarification
of the role and of the consequences
of such a factorization.

Having the factorization (\ref{cocody}) at their disposal,
Buslaev and Grecchi were able to realize
the
Hermitization (\ref{repreco})
via its $N-$step decomposition,
 \be
 H_{}\,\to\,H_{01}\,\to\,H_{02}\,\to\,\ldots\,\to\,H_{0N-1}\,
 \to\,\mathfrak{h}\,.
 \label{reaN}
 \ee
In their specific model these authors obtained
formula (\ref{reaN}) in which
$H=H^{(BG)}\equiv H_{00}$
represented a maximally computation-friendly and, in some sense,
maximally non-Hermitian Hamiltonian operator.
The strictly Hermitian final item
$\mathfrak{h}=\mathfrak{h}^{(BG)}\equiv H_{0N}$
proved reconstructed exactly, via auxiliary
isospectral-operator elements $H_{0j}=H_{0j}^{(BG)}$ of the
sequence with $0<j<N$. In their model they used a not too small
value of $N=7$ \cite{BG}.

In a model-independent setting
the existence of sequence (\ref{reaN}) is in fact a direct
consequence of the factorization (\ref{cocody}) of the Dyson map.
In what follows we will draw just multiple further consequences
out of these ansatzs. The underlying
factorization (\ref{cocody})
will be given here a
deeper algebraic
interpretation.
The recurrent $N-$step BGI
process (\ref{reaN})
of a
constructive guarantee of the unitarity of the evolution
will be characterized as a mere
special case of a much broader new class of
eligible
re-Hermitizations of Schr\"{o}dinger equation~(\ref{ostasta}).

\section{Two forms of re-Hermitization at $N=1$\label{druhasek}}


The current enormous growth of popularity of
the QHQM studies
only took place after the development
of certain robust DI
Hermitization of $H$
(see a concise summary of the history
in \cite{MZbook}).
A key characteristics of the process lies in the
amendment of the inner product in Hilbert space
(see, e.g.,
comprehensive reviews \cite{ali,Carl}).
One just keeps the Hamiltonian $H$ unchanged and
guarantees the unitarity of evolution
in a new,
physical Hilbert space ${\cal H}_{phys}$ obtained
by an amendment of the inner product
in the initial Hilbert space ${\cal H}_{math}$.

Given a candidate for the Hamiltonian $H \neq H^\dagger$
one cannot get rid of the feeling that
the BGI Hermitization (\ref{repreco})
looks more natural.
In what follows
we intend to strengthen the feeling. A constructive
analysis
will be provided in which
the standard BGI- and DI-path
Hermitizations
will be
shown competitive.
Moreover, we will
complement these options by a multitude of other
Hermitizations. We will show that
in the
QHQM theory with non-Hermitian Hamiltonians $H$
the optimal representation of the Hermitizable
Hamiltonians
may be selected
out of a much richer menu.

Formally, the equivalence between
the two alternative processes of the Hermitization
can be
characterized by the
diagram of review \cite{SIGMA},
 \be
 \label{krexhu}
  \ba
 \begin{array}{|c|}
 \hline
  {\cal H}_{math}
 \\
  \hline
 \ea
\
 \stackrel{{\rm DI\ path}}{ \longrightarrow }
 \
 \begin{array}{|c|}
 \hline
 {\cal H}_{phys} \\
  \hline
 \ea
\\ \ \ \
 \ \ \
\ \ \ \ \
  \stackrel{{\rm BGI\ path}}{ }
  \   \searrow
\ \ \ \ \  \nearrow \! \swarrow  \
  \stackrel{{\rm  equivalence\,. }\ }{ }
 \  \\ \ \
 \begin{array}{|c|}
  \hline
  \ \
 {\cal H}_{textbook} \ \  \\
  \hline
 \ea
 \\
\ea
 \ee
In spite of the fact that
the Hermiticity of $H$
only enters the scene ''in disguise'',
the calculations may be perceivably
facilitated when performed in ${\cal H}_{math}$
and interpreted along the DI or BGI path.
In comparison, a specific merit of the
technically more difficult
BGI Hermitization pattern
lies in the trivial inner product
in the target space ${\cal H}_{textbook}$,
rendering
the ultimate physical probabilistic interpretation
of the system in question maximally
transparent.


The first step towards our goal
(i.e., to establishing a deeper connection
between the BGI and DI Hermitizations) has been performed in
our recent
study \cite{EPJP}.
We complemented there the standard DI
methods of the one-step construction of the
physical-Hilbert-space metric $\Theta$
by an innovation
based on
an auxiliary $N-$term factorization of the operator,
 \be
 \Theta=\Theta_N=Z_NZ_{N-1} \ldots Z_2Z_1\,.
 \label{fakfa}
 \ee
Such an $N>1$ DI approach 
represented
the
quantum system in question
in an innovated
non-Hermitian Schr\"{o}dinger picture.
From our present perspective
the latter results 
can be perceived
as a preparatory step.  Its scope remained
interpreted as an amendment of the DI approach
in a way which
is more thoroughly discussed, at $N=2$ and at $N=3$,
in Appendix C.

In the present continuation of the project we will
point out that
one of the unnoticed but still most important
consequences of the metric-factorization postulate (\ref{fakfa})
has to be seen in
the parallel induction of the factorization of the
related Dyson map (\ref{cocody}).
This is our present main idea. Its use
will enable us to develop
a systematic
extension of both of the DI and BGI Hermitization patterns beyond
their existing realizations.

The core of our message will lie in
the shift of attention from the
factorized metric to the factorized Dyson map.
At $N=1$
this is closely connected with a
change of interpretation of the three vertices in
diagram
(\ref{krexhu}) and of the
Dyson-map-mediated relations
 \ben
 |\psi_n \pkt=\Omega\,|\psi_n \kt\,
 \een
which connect the
ket vectors
$|\psi_n \pkt \in {\cal H}_{textbook}$
and $ |\psi_n \kt \in  {\cal H}_{math}$.
This means that
diagram~(\ref{krexhu})
may be re-displayed in its more explicit version
 \be
 \label{rexhum}
  \ba
 \begin{array}{|c|}
 \hline
 {\rm  } |\psi_m\kt,
  \ {\rm  }
 \br \psi_n | \\ {\rm in\ }
  {\cal H}_{math}
 \\
  \hline
 \ea
\
 \stackrel{{\rm  metric \ }\Omega^\dagger\Omega}{ \longrightarrow }
 \
 \begin{array}{|c|}
 \hline
 {\rm  } |\psi_m\kt,
  \ {\rm  }
 \br \psi_n |\, \Omega^\dagger\Omega \\  {\rm in\ }
 {\cal H}_{phys} \\
  \hline
 \ea
\\  \ \ \ \ \ \
  \stackrel{{\rm Dyson\ map}}{ }
  \   \searrow
 \ \ \ \ \  \ \ \ \ \ \ \ \ \
  \nearrow \! \swarrow  \
  \stackrel{{\rm equivalence\,.}\ }{ }\ \ \ \ \
 \  \\
 \begin{array}{|c|}
 %
  \hline
 \Omega\, |\psi_m\kt,
  \
 \br \psi_n |\, \Omega^\dagger \\  {\rm in\ }
 {\cal H}_{textbook}  \\
  \hline
 \ea\ \
 \\
\ea
 \ee
Both the ket and bra vectors
are written down here in their preferred representation in ${\cal H}_{math}$.
Occasionally, the
``correct and physical'' Hilbert space
${\cal H}_{phys}$
will be re-denoted as ${\cal R}_{0}^{(0)}$.
Similarly, we will work with
${\cal H}_{math}$ {\it alias\,}
${\cal R}_{N}^{(0)}$
and with
${\cal H}_{textbook}$
{\it alias\,} ${\cal R}_{0}^{(N)}$.
Such an amendment of the notation conventions
will facilitate transition to the QHQM formalism using arbitrary $N$
in (\ref{cocody}).

\section{The $N=2$ BGI Hermitization and its two shortcuts\label{tretisek}}

In the studies
of various realistic non-Hermitian models
the references to the $N>1$ BGI-path
option (\ref{reaN}) remain
rare (see, e.g., \cite{Mateo,Tenney}).
One of the reasons is that
in contrast to the DI path
where one needs just
the inner-product metric $\Theta$,
the BGI-path option requires also the knowledge of
its ``square root'' Dyson map $\Omega$.
Another problem is that in some cases
the methodical gains provided by
the Hermitization $H \to \mathfrak{h} $ may
happen to be,
due the collateral technical complications,
weakened or even lost completely.
In such a case one simply has to return to the
traditional
calculations in ${\cal H}_{textbook}$.
Definitely, a more detailed exploration of the
BGI theory
is needed.

\subsection{The Hamiltonian-preserving DI shortcut}

In our preceding paper \cite{EPJP} it has been shown that
the decomposition (\ref{fakfa})
of the metric
into a product of $N$ factors
$Z_j$ can be associated 
with a sequence of certain specific,
auxiliary Hilbert spaces.
This has been shown to yield a specific
pattern
of a sequential DI-type
quasi-Hermitization of $H$.
In particular, after the first nontrivial choice 
of $N=2$
one has to consider
a triplet of 
spaces
 \be
 {\cal H}_{math}\to {\cal H}_{int}\to {\cal H}_{phys}\,.
 \label{nas}
 \ee
This means that
the three-Hilbert-space BGI-DI diagram (\ref{krexhu})
had to be replaced by the following
four-Hilbert-space upgrade
 \be
 \label{wd2humdi}
  \ba
  \ \ \ \
   \begin{array}{|c|}
 \hline
 {\cal H}_{math} \\
 \hline
 \ea
 \
 \stackrel{{\rm DI\ path,\ step\ a}}{ \longrightarrow }
 \
 \begin{array}{|c|}
 \hline
 {\cal H}_{int}^{}\\
 \hline
 \ea
 \
 \stackrel{{\rm DI\ path,\ step\ b}}{ \longrightarrow }
 \
 \begin{array}{|c|}
 \hline
 {\cal H}_{phys} \\
 \hline
 \ea
 \\  \ \ \ \ \ \ \ \ \ \ \ \ \ \
 \stackrel{{\rm  BGI\  path}}{ }    \searrow \ \ \ \ \
 \ \ \ \ \ \ \ \ \ \ \ \ \ \ \ \ \ \ \ \ \ \ \ \ \ \ \
 \ \ \ \ \  \nearrow \! \swarrow
 \stackrel{{\rm equivalence\,. }}{ }\ \ \
 \\ \ \ \ \ \
 \begin{array}{|c|}
 \hline
  \ \ \ \ \ \ \ \ \ \  \ \ \ \ \ \ \ \
  {\cal H}_{textbook}\ \ \ \ \ \ \ \ \ \ \ \ \ \ \ \ \ \
  \\
 \hline
 \ea
 \\
\ea
 \ee
In the upper line the diagram contains
the
two-step DI-path Hermitization
during which
one does not change
the Hamiltonian $H$
(nor the other
observables).
One just
has to keep the trivial metric assigned to  ${\cal H}_{math}$
and the full
factorized metric (\ref{fakfa}) assigned to
${\cal H}_{phys}$.

From the point of view of unitary quantum physics
it appeared useful to endow
the new, intermediate
Hilbert space ${\cal H}_{int}$
with a new and rather unusual inner-product
metric $Z_2$.
Although the latter metric 
has not been assigned any immediate physical meaning,
its introduction
was well motivated by mathematics.
For the purpose
(see also letter \cite{PLA} for a more detailed
explanatory commentary)
it was also recommended to assume
the Hermiticity and
positivity of the candidate
for the metric in ${\cal H}_{int}$.
Thus, the following,
Dyson-like product decomposition of
the auxiliary metric has been postulated,
 \be
 Z_2=\Omega_2^\dagger\,\Omega_2\,.
 \label{a[1]}
 \ee
Nevertheless, 
in contrast to our forthcoming considerations,
the existence of such a
re-factorization  was just based on a purely formal
analogy between the choices of $N=2$ and $N=1$.

\subsection{The Hamiltonian-amending alternative shortcut}

In our present paper, relation (\ref{a[1]})
will be reinterpreted as a reason for a
split of the three-term
sequence (\ref{nas}) into the two independent
and elementary
step-1 and step-2 subsequences,
 \be
 {\cal H}_{math}\to {\cal H}_{int}\,,\ \ \ \
 {\cal H}_{int}
 \to {\cal H}_{phys}\,.
 \label{gnas}
 \ee
In the left, step-1 doublet of spaces
the
target space ${\cal
H}_{int}$
can be perceived as a strict
analogue of the $N=1$ target space
${\cal H}_{phys}$ in diagram~(\ref{krexhu}). Similarly,
the initial-space item ${\cal H}_{int}$
in the right pair of spaces in (\ref{gnas})
can equally well be interpreted as
mimicking the initial $N=1$ space
${\cal H}_{math}$ of diagram (\ref{rexhum}).

This is our basic idea.
We will treat the two sub-steps (\ref{gnas}) separately.
Indeed,
in both of these cases
there exists no full-fledged
analogue of the textbook space ${\cal H}_{textbook}$
of Eq.~(\ref{rexhum}). This means,
strictly speaking,
that neither one of the $N=2$ analogues of
the textbook Hilbert space ${\cal H}_{textbook}$
could immediately
be assigned the status of a physical Hilbert space.
Still, in spite of this obstacle,
it is possible to preserve at least some of
the parallels with the $N=1$ scenario.


In the context of step 1,
${\cal H}_{math}\to{\cal H}_{int}$,
it is now easy to introduce a
manifestly unphysical formal analogue
of the ${\cal H}_{textbook}-$space component of diagram (\ref{rexhum})
and denote it, say, by symbol ${\cal K}_{math}$.
The related step-1/DI-path sub-Hermitization
process
may be then reinterpreted in terms of
the following three-Hilbert-space subdiagram
 \be
 \label{ukrexhum}
  \ba
  \ \ \ \ \ \ \
 \begin{array}{|c|}
 \hline
 {\rm  } |\psi_m\kt,
  \ {\rm  }
 \br \psi_n | \\ {\rm in\ }
  {\cal H}_{math}
 \\
  \hline
 \ea
\
 \stackrel{{\rm sub\!-\!Hermitization}}{ \longrightarrow }
 \
 \begin{array}{|c|}
 \hline
 {\rm  } |\psi_m\kt,
  \ {\rm  }
 \br \psi_n |\, \Omega^\dagger_2\Omega_2 \\  {\rm in\ }
 {\cal H}_{int} \\
  \hline
 \ea
\\
 \ \ \ \  \ \ \ \
  \stackrel{{\rm sub\!-\!map}\ \Omega_2}{ }
  \   \searrow
 \ \ \ \ \  \ \ \ \
 \ \ \ \
\ \ \ \ \  \nearrow \! \swarrow  \
  \stackrel{{\rm sub\!-\!equivalence.}\ }{ }
 \  \\
\ \
 \begin{array}{|c|}
  \hline
 \Omega_2\, |\psi_m\kt,
  \
 \br \psi_n |\, \Omega^\dagger_2 \\  {\rm in\ }
 {\cal K}_{math}  \\
  \hline
 \ea
 \\
\ea
 \ee
This is a perfect analogue of diagram (\ref{rexhum}).
The innovation
operates with the Dyson-map sub-factor $\Omega_2$
as defined in Eq.~(\ref{a[1]}).
What is more important
is that we need also a parallel transformation
of the Hamiltonian as well as of all of the other operators of
observables.
The initial non-Hermitian Hamiltonian operator $H$ acting in
${\cal H}_{math}$ and in ${\cal H}_{int}$ as well as in ${\cal H}_{phys}$
has to be replaced, in  ${\cal K}_{math}$, by
the new isospectral non-Hermitian Hamiltonian operator
$\Omega_2\,H\,\Omega_2^{-1}$
denoted, say, as $\mathfrak{h}_2$.
This enables us to finalize the
ultimate upgrade of diagram (\ref{ukrexhum}),
 \be
 \label{wukrexhum}
  \ba
  \ \ \ \ \ \ \
 \begin{array}{|c|}
 \hline
 {\rm  } |\psi_m\kt,
  \ {\rm  }
 \br \psi_n |  \\
 \ {\rm and\ } H\\ {\rm in\ }
  {\cal H}_{math}
 \\
  \hline
 \ea
\
 \stackrel{{\rm sub\!-\!Hermitization}}{ \longrightarrow }
 \
 \begin{array}{|c|}
 \hline
 {\rm  } |\psi_m\kt,
  \ {\rm  }
 \br \psi_n |\, \Omega^\dagger_2\Omega_2 \\
 \ {\rm and\ } H \\  {\rm in\ }
 {\cal H}_{int} \\
  \hline
 \ea
\\
 \ \ \ \  \ \ \ \
  \stackrel{{\rm sub\!-\!map}\ \Omega_2}{ }
  \   \searrow
 \ \ \ \ \  \ \ \ \
 \ \ \ \
\ \ \ \ \  \nearrow \! \swarrow  \
  \stackrel{{\rm sub\!-\!equivalence.}\ }{ }
 \  \\
\ \
 \begin{array}{|c|}
  \hline
 \Omega_2\, |\psi_m\kt,
  \
 \br \psi_n |\, \Omega^\dagger_2\\
 \ {\rm and\ } \mathfrak{h}_2=\Omega_2\,H\,\Omega_2^{-1} \\  {\rm in\ }
 {\cal K}_{math}  \\
  \hline
 \ea
 \\
\ea
 \ee
The
isospectrality relation connects now the initial
Hamiltonian $H$ with its new partner $\mathfrak{h}_2$.
The latter operator enters the respective
new hypothetical auxiliary
Schr\"{o}dinger equation of course.

A straightforward nature of the derivation of Eq.~(\ref{wukrexhum})
seems to encourage
the use
of the same factorization of our
positive definite ``generalized charge''
$Z_1$. Unfortunately,
such an attempt fails to work for
the reasons 
clarified in
Lemma Nr.~1 of paper \cite{EPJP}.
The correct and consistent analogue of relation
(\ref{a[1]}) had to have a more sophisticated 
inner-product-dependent
form
 \be
 Z_1=\Omega_1^{\ddagger (1)}\,\Omega_1\,,
 \ \ \ \ \Omega_1^{\ddagger (1)}=Z_2^{-1}\Omega_1^{\dagger}\,Z_2\,.
 \label{[15]}
 \ee
After the incorporation of this inner-product-adapted
factorization relation
the theory becomes consistent.
One
reveals an analogous three-Hilbert-space substructure
in the other, step-2/DI-path sub-Hermitization
process
in which the ket- and bra-vectors
as well as Hamiltonians may be given the explicit form
in their representation
in the
universal and unique working space ${\cal H}_{math}$,
 \be
 \label{fkpexhum}
  \ba
 \begin{array}{|c|}
 \hline
 {\rm  } |\psi_m\kt,
  \ {\rm  }
 \br \psi_n | Z_2 \\
 \ {\rm and\ } H
 \\
  {\rm in\ }
  {\cal H}_{int}
 \\
  \hline
 \ea
\
 \stackrel{{\rm sub\!-\!Hermitization}}{ \longrightarrow }
 \
 \begin{array}{|c|}
 \hline
 {\rm  } |\psi_m\kt,
  \ {\rm  }
 \br \psi_n |\,\Omega^\dagger_1 Z_2\Omega_1 \\
 \ {\rm and\ } H
 \\
  {\rm in\ }
 {\cal H}_{phys} \\
  \hline
 \ea
\\ \ \ \ \
  \stackrel{{\rm sub\!-\!map}\ \Omega_1}{ }
  \   \searrow
 \ \ \ \ \  \ \ \ \
 \ \ \ \
\ \ \ \ \  \nearrow \! \swarrow  \
  \stackrel{{\rm sub\!-\!equivalence.}\ }{ }\ \ \ \ \
 \  \\
 \begin{array}{|c|}
  \hline
 \Omega_1\, |\psi_m\kt,
  \
 \br \psi_n |\, \Omega^\dagger_1  Z_2 \\
 \ {\rm and\ } \mathfrak{h}_1=\Omega_1\,H\,\Omega_1^{-1}
 \\ {\rm in\ }
 {\cal K}_{phys}  \\
  \hline
 \ea
 \ \ \ \ \ \ \
 \\
\ea
 \ee
This also offers an explicit definition
of the respective operations of Hermitian conjugation
in ${\cal K}_{phys}$.
The corresponding Hamiltonian operator
acquires its
new and final ``textbook-compatible'' eligible,
BGI-shortcut form defined
by formula
$\mathfrak{h}_1=\Omega_1\,H\,\Omega_1^{-1}$.
This operator remains manifestly
non-Hermitian
in our unique manipulation Hilbert space ${\cal H}_{math}$
but, simultaneously, it can be perceived as safely
self-adjoint in the new physical Hilbert space ${\cal K}_{phys}$.
In this sense
one can conclude that
a new version of a quasi-Hermitian representation of the quantum
system in question
is obtained via
the partial, $\Omega_2-$mediated amendment of the metric, $I \to Z_2$,
accompanied by the partial, $\Omega_1-$mediated amendment of the
Hamiltonian itself, $H \to \mathfrak{h}_1$.

A qualitatively new mathematical feature of diagram (\ref{fkpexhum}) is that
in its upper-left corner the
operation of Hermitian conjugation is manifestly $Z_2$-dependent.
The sub-equivalence arrows then
extend the physical status from
${\cal H}_{phys}$ to
${\cal K}_{phys}$.
In the lower half of Eq.~(\ref{fkpexhum}) containing
the representation
of the new kets and bras in ${\cal K}_{phys}$
one may just add
the third eligible physical Hamiltonian operator
$\mathfrak{h}_1$.
The description of physics in the
new Hilbert space of states
${\cal K}_{phys}$ becomes equivalent
to the one described in its partners
${\cal H}_{phys}$ and ${\cal H}_{textbook}$.

The resulting extension
and completion of the considerations of Ref.~\cite{EPJP}
is new.
By
the replacement ${\cal K}_{math} \to {\cal K}_{phys}$
we managed to complement the
standard doublet
of the physical Hilbert spaces
${\cal H}_{textbook}$ and ${\cal H}_{phys}$
by
the third physical Hilbert space ${\cal K}_{phys}$.
In  this space
we
{\em simultaneously\,} amended {\em both\,}
the inner product metric
(via the operator replacement $I \to Z_2$)
{\em and\,} the
Hamiltonian (via the transformation $H \to \mathfrak{h}_1$).
Now, in only remains for us to add that
our analysis is also
opening,
as its methodical byproduct,
a new path towards a further, still more sophisticated
realization of the Hermitization programme.

\subsection{The ultimate six-Hilbert-space classification diagram at $N=2$}

At the moment, the realization of the programme is incomplete even
at $N=2$.
What remains to be done is
the search for the
missing link between the three Hilbert spaces
${\cal K}_{math}$,
${\cal K}_{phys}$ and
${\cal H}_{textbook}$.
The search
may use the same technique as above.
The initial, purely mathematical
Hilbert space ${\cal H}_{math}$
finds its analogue, in (\ref{wukrexhum}),
in the
unphysical
Hilbert space ${\cal K}_{math}$. Similarly,
the old, $N=1$ physical
Hilbert space ${\cal H}_{textbook}$
finds, in (\ref{fkpexhum}), its formal $N=2$ analogue
in the
new physical
Hilbert space ${\cal K}_{phys}$.
Thus,
the specification of the
connection between ${\cal K}_{math}$ and
${\cal K}_{phys}$
may still proceed by analogy.
Due to the
subtlety of the
inclusion of the
upper-left-space Hermiticity, the
completion of the analogy requires more care.
Still, its result is obvious and
has the form of diagram
 \be
 \label{izu2hum}
  \ba
 \begin{array}{|c|}
 \hline
 {\rm   }\Omega_2\,  |\psi_{m}\kt,
  \ {\rm   }
 \br \psi_{n} |\,\Omega_2^\dagger \\
  {\rm in\  }{\cal K}_{math}\\
 \hline
 \ea
 \
 \stackrel{{\rm amendment}}{ \longrightarrow }
 \
 \begin{array}{|c|}
 \hline
 {\rm  }\Omega_1\, |\psi_{m}\kt,
  \ {\rm   }
 \br \psi_{n} |\,\Omega_1^\dagger\, \Omega^\dagger_2\Omega_2 \\
  {\rm in\  }{\cal K}_{phys}\\
 \hline
 \ea
 \ \ \ \ \
 \\
 \stackrel{{\rm }\Omega_{21}}{ } \   \searrow \ \ \ \ \ \
 \ \ \ \ \
 \ \ \ \ \  \nearrow \! \swarrow  \ \ \
 \ \ \ \ \ \ \ \ \ \
   \\
 \begin{array}{|c|}
 \hline
 {\rm  }\Omega_2\,  \Omega_1\, |\psi_{m}\kt,
  \
 \br \psi_{n} |\,\Omega_1^\dagger\, \Omega_2^\dagger \\
  {\rm in\  }{\cal H}_{textbook}\\
 \hline
 \ea
 \ \ \ \ \ \ \ \
 \\
\ea
 \ee
where $\Omega_{21}=\Omega_2\,\Omega_1\,\Omega_2^{-1}$.
We can now
concatenate all of the three sub-diagrams
(\ref{ukrexhum}),
(\ref{fkpexhum}) and
(\ref{izu2hum}) into a single one.
This leads to
our final triangular six-space scheme at $N=2$,
 \be
 \label{zu2hum}
  \ba
   \begin{array}{|c|}
 \hline
 {\rm   } |\psi_{m}\kt,
  \ {\rm  }
 \br \psi_{n} | \\
 \ {\rm and\ } H
 \\
  {\rm in\  }{\cal H}_{math}\\
 \hline
 \ea
 \
 \stackrel{{\rm amendment}}{ \longrightarrow }
 \
 \begin{array}{|c|}
 \hline
 {\rm  } |\psi_{m}\kt,
  \ {\rm  }
 \br \psi_{n} |\, Z_2 \\
 \ {\rm and\ } H
 \\
  {\rm in\  }{\cal H}_{int}\\
 \hline
 \ea
 \
 \stackrel{{\rm amendment}}{ \longrightarrow }
 \
 \begin{array}{|c|}
 \hline
 {\rm   } |\psi_{m}\kt,
  \ {\rm   }
 \br \psi_{n} |\, Z_{2}\,Z_1 \\
 \ {\rm and\ } H
 \\
  {\rm in\  }{\cal H}_{phys}\\
 \hline
 \ea
 \\
 \stackrel{{\rm  }\Omega_2}{ } \   \searrow \ \  \ \ \ \ \
 \ \ \ \ \  \nearrow \! \swarrow  \
 \stackrel{{\rm }}{ }
 \ \ \ \ \ \ \ \ \
 \stackrel{{\rm  }\Omega_1}{ } \   \searrow \ \ \ \ \ \ \ \
 \ \ \ \ \  \nearrow \! \swarrow  \ \ \
 \\
 \begin{array}{|c|}
 \hline
 {\rm   }\Omega_2\,  |\psi_{m}\kt,
  \ {\rm   }
 \br \psi_{n} |\,\Omega_2^\dagger \\
 \ {\rm and\ } \mathfrak{h}_2=\Omega_2\,H\,\Omega_2^{-1}
 \\
  {\rm in\  }{\cal K}_{math}\\
 \hline
 \ea
 \
 \stackrel{{\rm amendment}}{ \longrightarrow }
 \
 \begin{array}{|c|}
 \hline
 {\rm  }\Omega_1\, |\psi_{m}\kt,
  \ {\rm   }
 \br \psi_{n} |\,\Omega_1^\dagger\, Z_2 \\
 \ {\rm and\ } \mathfrak{h}_1=\Omega_1\,H\,\Omega_1^{-1}
 \\
  {\rm in\  }{\cal K}_{phys}\\
 \hline
 \ea
 \\
 \stackrel{{\rm }\Omega_{21}}{ } \   \searrow \ \ \ \ \ \
 \ \ \ \ \  \nearrow \! \swarrow  \ \ \
   \\
 \begin{array}{|c|}
 \hline
 {\rm  }\Omega_2\,  \Omega_1\, |\psi_{m}\kt,
  \
 \br \psi_{n} |\,\Omega_1^\dagger\, \Omega_2^\dagger \\
 \ {\rm and\ } \mathfrak{h}=\Omega\,H\,\Omega^{-1}
 \\
  {\rm in\  }{\cal H}_{textbook}\\
 \hline
 \ea
 \\
\ea
 \ee
The diagram lists all of the relevant
Dyson
mappings, kets, bras and the
Hamiltonians represented in the single Hilbert space
${\cal H}_{math}$ {\it alias\,}
${\cal R}_{N}={\cal R}_{N}^{(0)}$.

Optionally, one could also expand here
$Z_2=\Omega_2^\dagger\,\Omega_2$ and,
in the light of Eq.~(\ref{[15]}),
$Z_{2}\,Z_1=\Omega^\dagger \,\Omega$
with $\Omega=\Omega_2\,\Omega_1$.
Such a correspondence
between the two-term factorization
of the metric and the two-term factorization
of the Dyson map
was already noticed in paper \cite{EPJP}.
Nevertheless,
this observation was only
interpreted there as a mathematical curiosity because
the subject of the paper, viz.,
the DI
Hermitization of $H$
did not require the construction of
the Dyson maps at all.

From our present perspective
the latter formulae look much more important
because
the knowledge of the
factorized Dyson maps is needed in (\ref{zu2hum}).
In an amended notation
with ${\cal K}_{phys}={\cal R}_{0}^{(1)}$
and ${\cal K}_{math}={\cal R}_{1}^{(1)}$
the resulting menu of the
four
eligible Hermitization strategies
may be now summarized
in the more compact $N=2$ flowchart
 \be
 \label{ewd2humdi}
  \ba
  \ \ \ \
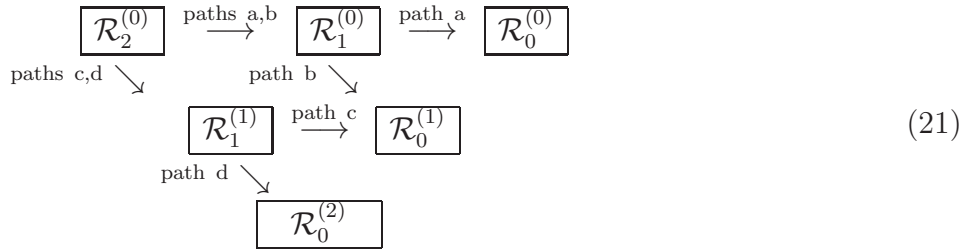
   \begin{array}{|c|}
 \hline
 {\cal R}_{2}^{(0)} \\
 \hline
 \ea
 \
 \stackrel{{\rm   paths \ a,b}}{ \longrightarrow }
 \
 \begin{array}{|c|}
 \hline
 {\cal R}_{1}^{(0)} \\
 \hline
 \ea
 \
 \stackrel{{\rm  path \ a}}{ \longrightarrow }
 \
 \begin{array}{|c|}
 \hline
 {\cal R}_{0}^{(0)}  \\
 \hline
 \ea
 \\
 \ \ \ \ \ \ \ \ \ \
 \stackrel{{\rm   paths \ c,d\ }}{ }    \searrow \ \
 \
 \
 \
 \ \ \ \ \
 \stackrel{{\rm   path \ b \ }}{ }    \searrow \
 \
 \ \ \ \ \ \ \ \ \ \ \
 \ \ \ \ \ \ \ \ \ \
 \ \ \ \ \ \ \ \ \ \
 \\
\ \ \ \
   \begin{array}{|c|}
 \hline
 {\cal R}_{1}^{(1)}  \\
 \hline
 \ea
 \
 \stackrel{{\rm  path \ c}}{ \longrightarrow }
 \
 \begin{array}{|c|}
 \hline
 {\cal R}_{0}^{(1)} \\
 \hline
 \ea
 \\
 \
 \stackrel{{\rm   path \ d\ }}{ }    \searrow \ \ \ \ \
 \ \
 \
 \ \ \ \ \ \ \ \ \ \
 \\ \ \ \
 \begin{array}{|c|}
 \hline
  \ \  {\cal R}_{0}^{(2)} \ \  \\
 \hline
 \ea
 \\
\ea
 \ee
For any preselected initial Hamiltonian $H$
this diagram offers the
choice among
the four, formally equally eligible options a, b, c and d.

\begin{table}[h]

\caption{The three optional choices of the
  model
  in QHQM at $N=2$.}
 \label{aas1x} \vspace{.4cm}
\centering
\begin{tabular}{||c||c|c|c||}
    \hline \hline
  Hermitization &BGI&new&DI\\
   (cf. diagram (\ref{ewd2humdi}))
    &(path d)&(paths b or c)&(path a)\\
\hline
 \hline
 physical metric& $I$&$Z_2$&$Z_2Z_1$\\
 Hamiltonian&$\mathfrak{h}$&$\mathfrak{h}_1$&$H$\\
 Hilbert space&${\cal R}_{0}^{(2)}$&${\cal R}_{0}^{(1)}$
 &${\cal R}_{0}^{(0)}$\\
\hline \hline
\end{tabular}
\end{table}

The list of the optional $N=2$ Hermitizations is now complete.
Presumably,
an ``optimal''
element selected out of
the four-path menu
(see also its rearrangement in Table \ref{aas1x})
will be, in any realistic application of the theory,
characterized by the minimalized costs of
the possible one- or two-step change of the inner product
(not needed along the BGI path~d)
and/or of the auxiliary transformation(s) of the Hamiltonian
(absent only along the DI path a).

The benefits of the separate
options will vary with the
input operator $H$
as well as with the feasibility
of the separate Hamiltonian-dependent Dyson sub-mappings.
A new feature
of the $N=2$ scheme
is that
the
choice of the path ending at ${\cal K}_{phys}$
is ambiguous. Its realization may
proceed
along the two alternative Hilbert-space triplets,
 \ben
 {\cal H}_{math}\to {\cal H}_{int}\to {\cal K}_{phys}\,,
 \, \ \ \ \ \
 {\cal H}_{math}\to {\cal K}_{math}\to {\cal K}_{phys}\,.
 \een
Both of these
paths
are, from the point of view of
Hermitization, equivalent.
In practice, therefore, one could often
speak just about the three distinct
Hermitizations as listed in
Table~\ref{aas1x}.

\section{The formalism using the factorizations with $N=3$\label{ctvrtasek}}

Without factorization
(i.e., at $N=1$)
one can distinguish just between the BGI Hermnitization
(working with the trivial inner product and changing $H$)
and the
DI $\Theta-$quasi-Hermnitization
(keeping the initially non-Hermitian Hamiltonian unchanged;
see diagram (\ref{krexhu})).
In section \ref{tretisek} we choose $N=2$ and we
revealed the existence of another, non-DI
quasi-Hermnitization
in which a new, non-BGI, partial amendment
$H \to \mathfrak{h}_1= \Omega_1\,H\,\Omega_1^{-1}$
of the Hamiltonian
had to be accompanied by a partial
simplification of the physical inner product,
$\Theta\ (\equiv Z_2\,Z_1)\, \to Z^2$.

In the $N=3$ diagram
 \be
 \label{zwd3humdi}
  \ba
  \\
   \begin{array}{|c|}
 \hline
 {\cal R}_3 \\
 \hline
 \ea
 \
 \stackrel{\rm DI\ path}{ \longrightarrow }
 \
 \begin{array}{|c|}
 \hline
 {\cal R}_{2}^{}\\
 \hline
 \ea
 \
 \
 \stackrel{\rm DI\ path}{ \longrightarrow }
 \
 \begin{array}{|c|}
 \hline
 {\cal R}_{1}^{}\\
 \hline
 \ea
 \
 \stackrel{\rm DI\ path}{ \longrightarrow }
 \
 \begin{array}{|c|}
 \hline
 {\cal R}_{0} \\
 \hline
 \ea
 \\  \ \ \ \ \  \ \ \ \ \ \ \
 \stackrel{{\rm BGI\ path }}{ }    \searrow \ \ \ \ \
 \ \ \ \ \ \ \  \ \ \ \ \   \ \ \ \ \   \ \ \ \ \ \ \ \ \
 \ \ \ \ \  \nearrow \! \swarrow
 \stackrel{{\rm equivalence\, }}{ }\ \ \ \ \ \ \
 \\
 \begin{array}{|c|}
 \hline
 \ \ \ \ \  \ \ \ \ \  \ \ \ \ \ \ \
  {\cal H}_{textbook}\ \ \ \ \ \ \ \ \ \  \ \ \ \ \ \ \    \\
 \hline
 \ea
 \\
\ea
 \ee
the meaning of the symbols ${\cal R}_k$ is obvious.
We only have to add here that
in full parallel to its $N=2$ predecessor in (\ref{wd2humdi}),
the DI-type upper-line Hermitization
pattern
is exceptional in the sense that
the
Hamiltonian remains the same along the whole
sequence of Hilbert spaces.
The related technical subtleties then
concern just the  step-by-step amendment of the metric.

Now we are going to offer a systematic description of
the non-DI-type options.
The new Hermitization paths
will be found
branched in a way which has not been noticed in our previous
papers \cite{EPJP,PLA}.
The basic idea behind
the ``branching''
will be shown again related to
the factorization of the metric.
Still, its description will
require a modification and extension of the paradigm.

The necessary change of the paradigm
will be counterintuitive, based on introduction of further
auxiliary inner-product spaces.
For these reasons, it makes sense to start its description
from its first nontrivial $N=3$ special case again.
In such a case one merely has to
replace the two $N=2$ doublets
of spaces in Eq.~(\ref{gnas}) by the three doublets,
 \be
 {\cal R}_{3}^{(0)}\to {\cal R}_{2}^{(0)}\,,\ \ \ \
 {\cal R}_{2}^{(0)}\to {\cal R}_{1}^{(0)}\,,\ \ \ \
 {\cal R}_{1}^{(0)}\to {\cal R}_{0}^{(0)}\,.
 \label{ggnas}
 \ee
After the three-term factorization
(\ref{fakfa}) of the correct metric making the ultimate
Hilbert space ${\cal H}_{phys}={\cal R}_0^{(0)}$ physical
we will assume again that all of the three
factors $Z_1$, $Z_2$ and $Z_3$
are
positive definite.
This opens the possibility of a
generalization of
the Dyson-map-resembling sub-factorization
of these factors.
We are now able to require
not only that we can always decompose
 \be
 Z_3=\Omega^\dagger_3\,\Omega_3
 \label{[27]}
 \ee
(relation reflecting the fact that operator $Z_3$ gets
factorized using a Dyson sub-map $\Omega_3$)
but also that we can always decompose
 \be
 Z_3\,Z_2=\widetilde{Z_2}=\widetilde{\Omega_2}^\dagger\,\widetilde{\Omega_2}\,,
 \ \ \ \  \widetilde{\Omega_2}=\Omega_3\,\Omega_2
 \label{[28]}
 \ee
(this decomposition determines the untilded Dyson sub-map $\Omega_2
=\Omega_3^{-1}\,\widetilde{\Omega_2}$) and that we can always decompose
 \be
 Z_3\,Z_2\,Z_1=\widetilde{Z_1}=\widetilde{\Omega_1}^\dagger\,\widetilde{\Omega_1}\,,
 \ \ \ \  \widetilde{\Omega_1}=\Omega_3\,\Omega_2\,\Omega_1
 \label{[29]}
 \ee
(this factorization
defines, finally, the third untilded Dyson sub-map $\Omega_1=
\widetilde{\Omega_2}^{-1}\,\widetilde{\Omega_1}$).

The most unexpected consequence
of the factorizations (\ref{[27]}),  (\ref{[28]}) and  (\ref{[29]})
is, according to
Lemma Nr.~2 of paper \cite{EPJP},
that they
specify, explicitly,
the
composite nature of the
conventional
Dyson map of Eq.~(\ref{cocody}),
 \be
 \Omega=\widetilde{\Omega_1}=\Omega_3\,\Omega_2\,\Omega_1\,.
 \label{[30]}
 \ee
Such a map connects
the initial, unphysical representation space ${\cal R}_{3}^{(0)}$
with
the conventional textbook physical Hilbert space ${\cal H}_{textbook}$.

Now we have to search for all of the possible
shortcuts of the $N-$step BGI Hermitizations,
i.e., for the multiple direct
and indirect analogues of the
ultimate DI path.
Once we return to the
$N=2$ cases for inspiration,
and once we recall
the two triangular sub-diagrams (\ref{ukrexhum})
and (\ref{fkpexhum})
we get encouraged to use
the same mathematics
yielding the $N=3$
triplet of the following concatenated triangles,
\be
 \label{f3humdi}
  \ba
     \begin{array}{|c|}
 \hline
 {\rm   } |\psi_{m}\kt,
  \ {\rm  }
 \br \psi_{n} | \\
  {\rm in\ }{\cal R}_{3}^{(0)}\\
 \hline
 \ea
 \
 \stackrel{{\rm }Z_3}{ \longrightarrow }
 \
 \begin{array}{|c|}
 \hline
 {\rm  } |\psi_{m}\kt,
  \ {\rm  }
 \br \psi_{n} |\, Z_3 \\
 {\rm in\ }{\cal R}_{2}^{(0)}\\
 \hline
 \ea
 \
 \stackrel{{\rm }Z_2}{ \longrightarrow }
 \
 \begin{array}{|c|}
 \hline
 {\rm   } |\psi_{m}\kt,
  \ {\rm   }
 \br \psi_{n} |\, Z_3\,Z_2 \\
 {\rm in\ }{\cal R}_{1}^{(0)}\\
 \hline
 \ea
 \
 \stackrel{{\rm }Z_1}{ \longrightarrow }
 \
 \begin{array}{|c|}
 \hline
 {\rm   } |\psi_{m}\kt,
  \ {\rm   }
 \br \psi_{n} |\, Z_3\,Z_2\,Z_1 \\
 {\rm in\ }{\cal R}_{0}^{(0)}\\
 \hline
 \ea
  \\
  \stackrel{{\rm  }\Omega_3}{ } \   \searrow \ \  \ \ \
  \ \ \ \
 \nearrow \! \swarrow  \
 \stackrel{{\rm }}{ }
 \ \ \ \
 \stackrel{{\rm  }\Omega_2}{ } \   \searrow \ \ \ \ \ \ \
 \ \ \
  \nearrow \! \swarrow
 \ \ \  \ \ \ \ \ \ \
 \stackrel{{\rm  }\Omega_1}{ } \   \searrow \ \ \ \ \ \ \ \
 \ \ \
 \nearrow \! \swarrow
  \ \ \ \ \ \ \ \ \ \
  \\
     \begin{array}{|c|}
 \hline
 \Omega_3\, |\psi_{m}\kt,
  \
 \br \psi_{n} |\,\Omega_3^\dagger \\
 {\rm in\ }{\cal J}_{2}\\
 \hline
 \ea
 \ \ \ \
 \
 \begin{array}{|c|}
 \hline
 \Omega_2\, |\psi_{m}\kt,
  \
 \br \psi_{n} |\,\Omega_2^\dagger\, Z_3 \\
 {\rm in\ }{\cal J}_{1}\\
 \hline
 \ea
 \ \ \ \
 \
 \begin{array}{|c|}
 \hline
 \Omega_1\, |\psi_{m}\kt,
  \
 \br \psi_{n} |\,\Omega_1^\dagger\,\,  Z_3\,Z_2 \\
 {\rm in\ }{\cal J}_{0}\\
 \hline
 \ea\,.
 \ea
 \ee
In this diagram the arrows at
$\Omega_k$
mark
the transitions from the initial spaces
${\cal R}_{k}^{(0)}$ to a new triplet
${\cal J}_{k-1}$ with $k=3, 2$ and $1$.
These changes of Hilbert spaces are
accompanied by the respective replacements of
$H$ by the respective
new isospectral Hamiltonian
operators $\mathfrak{h}_k= \Omega_k\,H\,\Omega_k^{-1}$.
In a formal parallel with Eq.~(\ref{fkpexhum}) (where we had
$N=2$)
we now arrive at the
ultimate physical $N=3$ DI-path Hilbert space
${\cal R}_0^{(0)}$ as well as
at its new, manifestly physical
partner ${\cal J}_0$.





In Eq.~(\ref{f3humdi})
the unphysical
intermediate Hamiltonians
$\mathfrak{h}_2$ and
$\mathfrak{h}_3$
may happen to be available and tractable as easily as
the original
input Hamiltonian $H$.
As we already mentioned,
it may make sense to
use of such a knowledge
and to study the unusual Hermitizations
passing through these two less usual intermediates.
Nevertheless,
as long as we have $N=3$, there still exists
a methodically challenging missing-line gap
between
the triplet of
spaces ${\cal J}_k$ in~(\ref{f3humdi})
and the ultimate unique textbook Hilbert space
${\cal H}_{textbook}$.

We already explained above how
such a gap has to be filled.
Leaving, therefore, the details of the
construction to the readers
let us merely formulate the final result.
It states that the role of the $N=3$
analogue of the six-space $N=2$ triangle
(\ref{zu2hum})
is now played by the following ten-space triangular lattice
 \be
 \label{3humdi}
  \ba
  \\
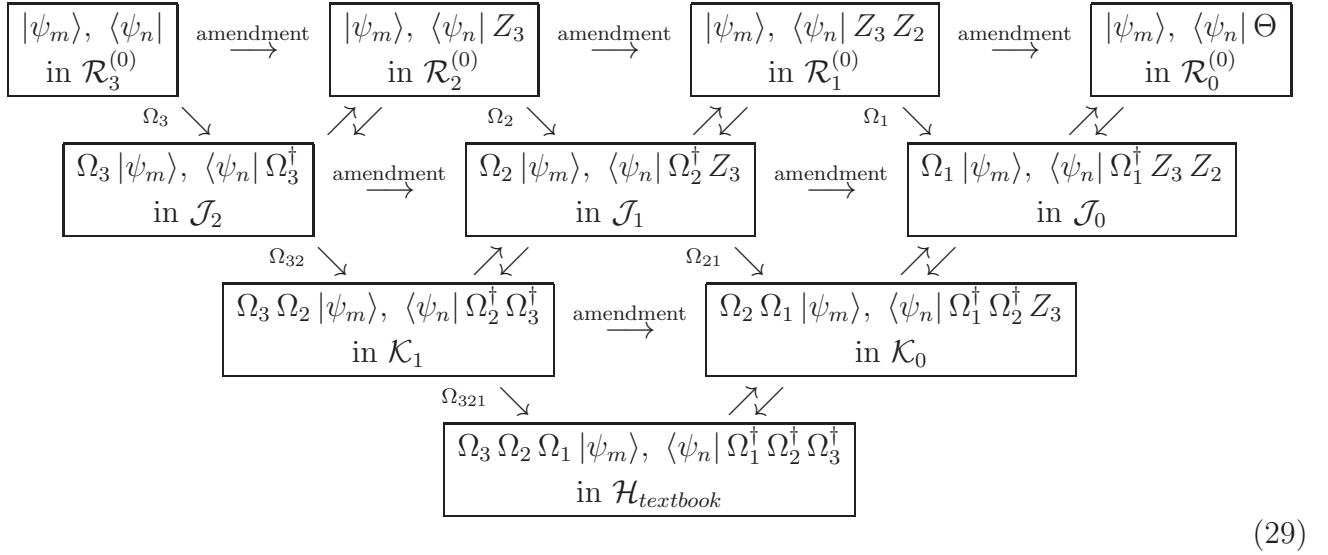
     \begin{array}{|c|}
 \hline
 {\rm   } |\psi_{m}\kt,
  \ {\rm  }
 \br \psi_{n} | \\
  {\rm in\ }{\cal R}_{3}^{(0)}\\
 \hline
 \ea
 \
 \stackrel{{\rm amendment}}{ \longrightarrow }
 \
 \begin{array}{|c|}
 \hline
 {\rm  } |\psi_{m}\kt,
  \ {\rm  }
 \br \psi_{n} |\, Z_3 \\
 {\rm in\ }{\cal R}_{2}^{(0)}\\
 \hline
 \ea
 \
 \stackrel{{\rm amendment}}{ \longrightarrow }
 \
 \begin{array}{|c|}
 \hline
 {\rm   } |\psi_{m}\kt,
  \ {\rm   }
 \br \psi_{n} |\, Z_3\,Z_2 \\
 {\rm in\ }{\cal R}_{1}^{(0)}\\
 \hline
 \ea
 \
 \stackrel{{\rm amendment}}{ \longrightarrow }
 \
 \begin{array}{|c|}
 \hline
 {\rm   } |\psi_{m}\kt,
  \ {\rm   }
 \br \psi_{n} |\, \Theta_{} \\
 {\rm in\ }{\cal R}_{0}^{(0)}\\
 \hline
 \ea
  \\
  \stackrel{{\rm  }\Omega_3}{ } \   \searrow \ \  \ \ \ \
 \ \ \ \ \  \nearrow \! \swarrow  \
 \stackrel{{\rm }}{ }
 \ \ \ \ \ \ \ \ \
 \stackrel{{\rm  }\Omega_2}{ } \   \searrow \ \ \ \ \ \ \ \
 \ \ \ \ \  \nearrow \! \swarrow  \ \ \  \ \ \ \  \ \ \ \
 \stackrel{{\rm  }\Omega_1}{ } \   \searrow \ \ \ \ \ \ \ \
 \ \ \ \ \  \nearrow \! \swarrow  \ \ \
  \\
     \begin{array}{|c|}
 \hline
 \Omega_3\, |\psi_{m}\kt,
  \
 \br \psi_{n} |\,\Omega_3^\dagger \\
 {\rm in\ }{\cal J}_{2}\\
 \hline
 \ea
 \
 \stackrel{{\rm amendment}}{ \longrightarrow }
 \
 \begin{array}{|c|}
 \hline
 \Omega_2\, |\psi_{m}\kt,
  \
 \br \psi_{n} |\,\Omega_2^\dagger\, Z_3 \\
 {\rm in\ }{\cal J}_{1}\\
 \hline
 \ea
 \
 \stackrel{{\rm amendment}}{ \longrightarrow }
 \
 \begin{array}{|c|}
 \hline
 \Omega_1\, |\psi_{m}\kt,
  \
 \br \psi_{n} |\,\Omega_1^\dagger\, Z_{3}\, Z_{2} \\
 {\rm in\ }{\cal J}_{0}\\
 \hline
 \ea
  \\
  \stackrel{{\rm  }\Omega_{32}}{ } \   \searrow \ \  \ \ \ \ \
 \ \ \ \ \  \nearrow \! \swarrow  \
 \stackrel{{\rm }}{ }
 \ \ \ \ \ \ \ \ \ \ \ \ \ \
 \stackrel{{\rm  }\Omega_{21}}{ } \   \searrow \ \ \ \ \ \ \ \
 \ \ \ \ \  \nearrow \! \swarrow  \ \ \ \ \ \ \
  \\
  \begin{array}{|c|}
 \hline
 \Omega_3\, \Omega_2\,  |\psi_{m}\kt,
  \
 \br \psi_{n} |\,\Omega_2^\dagger\,\Omega_3^\dagger  \\
 {\rm in\ }{\cal K}_{1}\\
 \hline
 \ea
 \
 \stackrel{{\rm amendment}}{ \longrightarrow }
 \
 \begin{array}{|c|}
 \hline
 \Omega_2\, \Omega_1\,  |\psi_{m}\kt,
  \
 \br \psi_{n} |\,\Omega_1^\dagger\,\Omega_2^\dagger \, Z_{3} \\
 {\rm in\ }{\cal K}_{0}\\
 \hline
 \ea
 \\
 \stackrel{{\rm }\Omega_{321}}{ } \   \searrow \ \ \ \ \ \ \ \ \ \ \ \ \ \
 \ \ \ \ \  \nearrow \! \swarrow  \ \ \ \ \ \ \
   \\
 \begin{array}{|c|}
 \hline
 \Omega_3\, \Omega_2\,  \Omega_1\, |\psi_{m}\kt,
  \ {\rm   }
 \br \psi_{n} |\,\Omega_1^\dagger\, \Omega_2^\dagger \, \Omega_3^\dagger\\
 {\rm in\ }{\cal H}_{textbook} \\
 \hline
 \ea
 \\
\ea
 \ee
The abbreviations
$\Omega_{32}=\Omega_3\,\Omega_2\,\Omega_3^{-1}$,
$\Omega_{21}=\Omega_2\,\Omega_1\,\Omega_2^{-1}$
and
$\Omega_{321}=\Omega_3\,\Omega_{21}\,\Omega_3^{-1}$
have a recurrent structure. This reflects the fact that
under the three-term factorization assumption (\ref{[30]})
we had to introduce the
two new Hilbert spaces denoted as
${\cal K}_1$ and ${\cal K}_0$.
In these spaces
the Hamiltonians may easily be found to have the respective
forms
$\mathfrak{h}_{32}=\Omega_3\,\Omega_2\,H\,\Omega_2^{-1}\,\Omega_3^{-1}$
(which is still non-Hermitian in ${\cal K}_1$) and
$\mathfrak{h}_{21}=\Omega_2\,\Omega_1\,H\,\Omega_1^{-1}\,\Omega_2^{-1}$
(which is quasi-Hermitian in~${\cal K}_0$).

\begin{table}[h]

\caption{The BGI Hermitization
and its three quasi-Hermitian shortcuts at $N=3$. The
definitions of
$\mathfrak{h}_{j}=\Omega_{j}\,H\,\Omega_{j}^{-1}$,
$\mathfrak{h}_{jj-1}=\Omega_{jj-1}\,\mathfrak{h}_{j}\,\Omega_{jj-1}^{-1}$
and of
$\mathfrak{h}_{jj-1j-2}=\Omega_{jj-1j-2}\,
\mathfrak{h}_{jj-1}\,\Omega_{jj-1j-2}^{-1}$ are recurrent.}
 \label{bbs1x} \vspace{.04cm}
\centering
\begin{tabular}{||c||c|c|c||}
    \hline \hline
   \multicolumn{1}{||c||}{BGI} &
    \multicolumn{3}{c||}{quasi-Hermiticity shortcuts}\\
   \hline
 \multicolumn{1}{||c||}{} &the $1^{\rm st}$&the $2^{\rm nd}$
 &the $3^{\rm rd}$ (DI)\\
    \hline \hline
 \multicolumn{4}{||c||}{non-Hermitian mathematical input}\\
    \hline
    \hline
    $H$&&&\\
    $\mathfrak{h}_3$&$H$&&\\
    $\mathfrak{h}_{32}$&$\mathfrak{h}_{2}$&$H$&\\
    \hline \hline
    \multicolumn{4}{||c||}{physical versions of Hamiltonians}\\
    \hline
    \hline
   $\mathfrak{h}_{321}$&$\mathfrak{h}_{21}$&$\mathfrak{h}_1$&$H$\\
    \hline \hline
    \multicolumn{4}{||c||}{physical Hilbert-space metrics}\\
    \hline
 $I$&$Z_3$&$Z_3Z_2$&$Z_3Z_2Z_1$\\
\hline \hline
\end{tabular}
\end{table}

We see that the
$N=3$ version of the formalism may be characterized by the
menu of as many as four alternative physical scenarios
realized in
as many as four physical Hilbert spaces
${\cal R}_0^{(0)}$, ${\cal J}_0$
({\it alias\,}
${\cal R}_0^{(1)}$), ${\cal K}_0$
({\it alias\,}
${\cal R}_0^{(2)}$), and ${\cal H}_{textbook}$
({\it alias\,},
${\cal R}_0^{(3)}$).
In a different format this observation
is also summarized in Table~\ref{bbs1x}.
In practice, certainly, the choice
out of this menu
will critically depend on the
user-friendliness (i.e., on the complexity)
of the respective Schr\"{o}dinger equations
and/or on the accessibility of
the desirable phenomenological predictions,
i.e., on the feasibility of the evaluation
of certain relevant matrix elements.

\section{The classification of the quasi-Hermitian
reformulations of QHQM at any $N$}

Our detailed derivation of the
latter, $N=3$ diagram (\ref{3humdi})
can be read as a methodical guide.
At arbitrary $N>3$ the process of its generalization
may start again
from the Hamiltonian-preserving,
metric-mediated DI-type
Hermitization which
proceeds through an $(N+1)-$plet of
Hilbert spaces
 \be
 {\cal R}_{N}^{(0)}\to{\cal R}_{N-1}^{(0)}\to\ldots\to{\cal R}_{1}^{(0)}\to
 {\cal R}_{0}^{(0)}\,.
 \label{usmultpl}
 \ee
The leftmost
symbol denotes here again the
initial, preferred representation space ${\cal R}_{N}^{(0)}={\cal H}_{math}$
with the inner-product metric equal to identity.
At the right-hand side of the set we have the ultimate physical space
${\cal R}_{0}^{(0)}={\cal H}_{phys}$
in which the Hamiltonian
$H$ becomes
self-adjoint, the quantum
measurements become
correctly interpreted as probabilistic,
and the unitarity of the evolution becomes guaranteed by
the Stone theorem.

In this direction we already made the first steps
in \cite{EPJP} where we managed to
augment, in the language of diagrams, the two-line DI
pattern (\ref{zwd3humdi}) from $N=3$ to any $N$,
 \be
 \label{uzwedli}
  \ba
  \\
   \begin{array}{|c|}
 \hline
 {\cal R}_N^{(0)} \\
 \hline
 \ea
 \
 \stackrel{\rm DI\ step\ a}{ \longrightarrow }
 \
 \begin{array}{|c|}
 \hline
 {\cal R}_{N-1}^{(0)}\\
 \hline
 \ea
 \
 \
 \stackrel{\rm DI\ step\ b}{ \longrightarrow }
 \
 \ldots
 \
 \stackrel{\rm DI\ step\ y}{ \longrightarrow }
 \
 \begin{array}{|c|}
 \hline
 {\cal R}_{1}^{(0)}\\
 \hline
 \ea
 \
 \stackrel{\rm DI\ step\ z}{ \longrightarrow }
 \
 \begin{array}{|c|}
 \hline
 {\cal R}_{0}^{(0)} \\
 \hline
 \ea
 \\  \ \ \ \ \  \ \ \ \ \ \ \
 \stackrel{{\rm BGI\ path}}{ }    \searrow \ \ \ \ \
 \ \ \ \ \ \ \  \ \ \ \ \   \ \ \ \ \   \ \ \ \ \ \ \ \ \
 \ \ \ \ \  \ \ \ \ \  \ \ \ \ \ \ \ \  \ \ \ \
 \ \ \ \ \ \ \
 \ \ \ \ \  \nearrow \! \swarrow
 \stackrel{{\rm equivalence\,. }}{ }\ \ \ \ \ \ \
 \\
 \begin{array}{|c|}
 \hline
 \ \ \ \ \  \ \ \ \ \  \  \ \ \ \
 \ \ \ \ \  \ \ \ \ \  \ \ \ \ \ \ \
  {\cal H}_{textbook}\ \ \ \ \ \ \ \ \ \  \ \
 \ \ \ \ \  \ \ \ \ \  \ \ \ \ \ \ \  \\
 \hline
 \ea
 \\
\ea
 \ee
In place of the single intermediate Hilbert space ${\cal H}_{int}$
of Eq.~(\ref{wd2humdi}) at $N=2$,
the general scheme comprises
as many as $N-1$ different, zero-superscripted intermediate
Hilbert spaces. In this sense
the main
result of our preceding paper~\cite{EPJP} may be characterized as
a discovery of a
one-to-one correspondence between the separate Hilbert-space elements of
the $(N+1)-$plet (\ref{usmultpl})
and
the
individual metric-operator factors $Z_j$ (cf. our present
Eq.~(\ref{fakfa}) above and the
Tables Nr. 1 and 2
and the equation Nr. (16) in {\it loc. cit}).

In our present paper
we just promised to generalize the latter scheme.
In this spirit we started from
the triangular
three-space diagram (\ref{krexhu}) at $N=1$,
and we obtained
the six-space triangular diagram (\ref{zu2hum}) at $N=2$
and the analogous ten-space diagram (\ref{3humdi}) at $N=3$.
Thus, the climax of the story may be expected to provide, at any $N$, a
triangular diagram which will
comprise
as many as
 {\tiny
 $\!\!\begin{array}{c}\!\!\left (\!\!\begin{array}{c}N+2\\2
 \end{array}\!\!\right )
 \end{array}\!\!\!$
 }
Hilbert spaces.

Our derivation
of the extrapolation
may proceed quickly.
We may start again from the known
DI-type $N-$step recipe of paper
\cite{EPJP}.
With the ultimate physical metric
$\Theta$ given by formula (\ref{fakfa}) we may characterize
the starting step by the diagram
 \ben
  \ba
   \begin{array}{|c|}
 \hline
 {\rm   } |\psi_{m}\kt,
 \br \psi_{n} | \\
 {\rm for }\ {\cal R}_{N}^{(0)}\\
 \hline
 \ea
 \
 \stackrel{\rm step\ 1}{ \longrightarrow }
 \
 \begin{array}{|c|}
 \hline
 {\rm  } |\psi_{m}\kt,
 \br \psi_{n} |Z_N \\
 {\rm for }\ {\cal R}_{N-1}^{(0)}\\
 \hline
 \ea
 \
 \stackrel{\rm step\ 2}{ \longrightarrow }
 \
 \begin{array}{|c|}
 \hline
 {\rm  } |\psi_{m}\kt,
 \br \psi_{n} |Z_N Z_{N-1} \\
 {\rm for }\ {\cal R}_{N-2}^{(0)}\\
 \hline
 \ea
 \
 \stackrel{\rm step\ 3 }{ \longrightarrow }
 \ \ldots
 \
 \stackrel{\rm step\ N }{ \longrightarrow }
 \
 \begin{array}{|c|}
 \hline
 {\rm   } |\psi_{m}\kt,
 \br \psi_{n} |\Theta\\
 {\rm for }\ {\cal R}_{0}^{(0)}\\
 \hline
 \ea\,.
\ea
 \een
The
mathematics behind such a structure
has been fully developed elsewhere \cite{EPJP,EPL}:
The diagram just
generalizes
its two special cases as listed
in Appendix C (cf. Eqs. (\ref{u2hum}) and (\ref{ff3humdi}) below).
Thus, we may
proceed along the same lines as above.
The general construction
will be fully analogous to
the preceding special cases so that the
description of
the details may be left to the readers.
Let us
display just the final result which generalizes
diagram (\ref{ewd2humdi}) beyond $N=2$,
 \be
 \label{gewd2hu}
  \ba
  \ \ \ \
   \begin{array}{|c|}
 \hline
 {\cal R}_{N}^{(0)} \\
 \hline
 \ea
 \
 \stackrel{{\rm  2^{N-1}\ paths  }}{ \longrightarrow }
 \
 \begin{array}{|c|}
 \hline
 {\cal R}_{N-1}^{(0)}\\
 \hline
 \ea
 \
 \stackrel{{\rm 2^{N-2}\ paths  }}{ \longrightarrow }
 \
 \begin{array}{|c|}
 \hline
 {\cal R}_{N-2}^{(0)}\\
 \hline
 \ea
 \
 \stackrel{{\rm  }}{ \longrightarrow }
 \
 \ldots
 \
 \stackrel{{\rm 2\ paths }}{ \longrightarrow }
 \
 \begin{array}{|c|}
 \hline
 {\cal R}_{1}^{(0)}\\
 \hline
 \ea
 \
 \stackrel{{\rm 1\ path }}{ \longrightarrow }
 \
 \begin{array}{|c|}
 \hline
 {\cal R}_{0}^{(0)} \\
 \hline
 \ea
 \\
 \stackrel{{\rm   2^{N-1}\ paths  }}{ }    \searrow \ \
 \
 \
 \
 \ \ \ \ \
 \stackrel{{\rm  2^{N-2}\ paths   }}{ }    \searrow \ \
 \
 \
 \
 \ \ \ \
 \ \ \ \
 \
 \ \ \ \ \
 \stackrel{{\rm    }}{ }    \searrow \ \
 \
 \
 \
 \ \ \ \
 \
 \ldots
 \ \ \ \ \
 \ \ \ \ \
 \stackrel{{\rm  1\ path }}{ }    \searrow \
 \
 \ \ \ \ \
 \ \ \ \ \
 \ \ \ \ \
 \\
 \ \ \ \ \
   \begin{array}{|c|}
 \hline
 {\cal R}_{N-1}^{(1)} \\
 \hline
 \ea
 \
 \stackrel{{\rm  2^{N-2}\ paths  }}{ \longrightarrow }
 \
 \begin{array}{|c|}
 \hline
 {\cal R}_{N-2}^{(1)}\\
 \hline
 \ea
 \
 \stackrel{{\rm  }}{ \longrightarrow }
 \
 \ \
 \ldots
 \ \
 \
 \stackrel{{\rm  }}{ \longrightarrow }
 \
 \begin{array}{|c|}
 \hline
 {\cal R}_{1}^{(1)}\\
 \hline
 \ea
 \
 \stackrel{{\rm N-1\ paths }}{ \longrightarrow }
 \
 \begin{array}{|c|}
 \hline
 {\cal R}_{0}^{(1)} \\
 \hline
 \ea
 \\
 \stackrel{{\rm  2^{N-2}\ paths   }}{ }    \searrow \ \
 \
 \ \ \ \
 \ \ \ \ \
 \
 \ \ \ \ \ \
 \stackrel{{\rm   }}{ }    \searrow \ \
 \ \ \ \
 \
 \
 \ \
 \
 \ldots
 \
 \ \ \ \ \
 \ \ \ \ \
 \
 \stackrel{{\rm   }}{ }    \searrow \
 \
 \ \ \ \
 \ \ \ \ \ \ \ \ \ \
 \\
 \ldots
\ \ \ \ \ \
 \ldots
\ \ \ \ \ \
 \ldots
\ \ \ \ \ \
 \ldots
\ \ \ \ \ \
 \ldots
\ \ \ \ \ \
 \ldots\ \ \ \ \ \
 \ldots
 \\ \
 \stackrel{{\rm  2\ paths }}{ }    \searrow \ \
 \
 \
 \ \
 \stackrel{{\rm N-1\  paths }}{ }    \searrow \ \
 \
 \
 \
 \ \ \ \
 \ \ \ \ \ \ \ \ \ \
\ \ \ \
\\
   \begin{array}{|c|}
 \hline
 {\cal R}_{1}^{(N-1)}  \\
 \hline
 \ea
 \
 \stackrel{{\rm 1\ path}}{ \longrightarrow }
 \
 \begin{array}{|c|}
 \hline
 {\cal R}_{0}^{(N-1)} \\
 \hline
 \ea
 \\
 \
 \stackrel{{\rm 1\  path }}{ }    \searrow \ \ \ \ \
 \ \
 \
 \ \ \ \ \ \ \ \ \ \
 \\ \ \ \
 \begin{array}{|c|}
 \hline
  \ \ {\cal R}_{0}^{(N)} \ \  \\
 \hline
 \ea\,.
 \\
\ea
 \ee
For the enumeration of all of the paths of the eligible Hermitizations,
is is now sufficient to
fix one of integers $k=0,1,\ldots,N$
and
the Hilbert-space sequence
 \be
 {\cal R}_N^{(0)}\, \to \,
 {\cal R}_{N-1}^{(k_1)}\, \to \, \ldots \,\to\,
 {\cal R}_1^{(k_{N-1})}\, \to \,
 {\cal R}_0^{(k)}
 \label{smula}
 \ee
in which the superscripts are variable.

\begin{lemma}
\label{lemmaone}
Every eligible Hermitization path (\ref{smula})
is determined by its set of superscripts such that
 \be
 0=k_0\leq k_1\leq k_2\leq \ldots \leq k_{N-1}\leq k_N=k\,.
 \label{smulab}
 \ee
\end{lemma}

The number of all of the
eligible Hermitization paths is large, equal to $2^N$.
Nevertheless, only the
Hilbert spaces ${\cal R}_0^{(k)}$ with $k=0,1,\ldots,N$
offer a consistent Hermitian or
quasi-Hermitian picture of physics.
This enables us to
compactify
our ultimate
classification of the (quasi-)Hermitizations
in the form of Table~\ref{zxs1x}.
.

\begin{table}[h]

\caption{The complete list of possible Hermitizations of a given
non-Hermitian Hamiltonian $H$.}
 \label{zxs1x} \vspace{.4cm}
\centering
\begin{tabular}{||c||lllll|c|c||c||}
    \hline \hline
 &
\multicolumn{5}{c|}{non-Hermitian Hamiltonians}& physical &
\multicolumn{1}{c||}{the amended metric}&the $H \to H_{k N}$\\
$k$ &\multicolumn{5}{c|}{(auxiliary sequence)}&Hamiltonian&
\multicolumn{1}{c||}{in Hilbert space}&transformation\\
\hline \hline
 0&$H_{}$&$H_{01}$&$H_{02}$&$\ldots$&$H_{0N-1}$&$\mathfrak{h}$&$I$&in
 $N$ steps (BGI)\\
 1&&$\ H_{}$&$H_{12}$&$\ldots$&$H_{1N-1}$&$H_{1N}$&$Z_N$&in $N-1$ steps\\
$\vdots$&& & $\ddots$ & $\ddots$ & $\ \ \ \vdots$& $\ \ \ \vdots$
& $\ \ \vdots$& $\vdots$
 \\
$N-2$&&&&$H_{}$&$H_{N-2N-1}$&$H_{N-2N}$&$Z_NZ_{N-1}\ldots Z_3$&in two steps\\
$N-1$&&&&&$\ \ \ H_{}$&$H_{N-1N}$&$Z_NZ_{N-1}\ldots Z_3Z_2$&in one step\\
$N$&&&&&&$H_{NN}\equiv H$&$Z_NZ_{N-1}\ldots Z_3Z_2Z_1$&no change (DI)\\
\hline \hline
\end{tabular}
\end{table}



According to this ultimate
classification
the role of the correct physical Hamiltonian
can be played,
at any $N$, either by the
Hermitian operator $\mathfrak{h}=H_{0N}$
or, at the $k-$th non-trivial inner-product metric,
by one of the
quasi-Hermitian operators $H_{kN}\,$ with $k=1,2,\ldots,N$.
What is not displayed in
Table \ref{zxs1x} are the explicit
formulae which would specify
the Hamiltonians at all of the admissible
values of the subscripts.
The
derivation of these formulae
is easy and, moreover, it may be
further facilitated
by the straightforward extrapolation of the $N=2$ formulae of
section \ref{tretisek} (cf. also Table~\ref{aas1x})
and of their $N=3$ descendants in section \ref{ctvrtasek}
(cf. Table \ref{bbs1x}).
Hence, the task is left to the readers.


\section{The path-dependence of complexity in
a schematic three-level quantum model\label{sestasek}}

During the future use
of our
formalism in quantum physics of (hiddenly) unitary systems,
one of the key decisions to be made will concern
the
choice of the
physical space
${\cal R}_0^{(k)}$,
i.e., the selection of the
$k-$th-line-strategy out of
Table \ref{zxs1x}.
The decision
may be expected to
vary with the dynamical
information encoded in the input operator $H$.
For this reason
there probably exists no universally valid
recommendation of one of
the alternatives. Still,
let us now return, once more, to the
``first nontrivial''
$N=3$ scenario, and let us try to use
such a special case
for a
more concrete illustration and
a sampling
of at least some of the
meaningful criteria.

At $N=3$
the general menu of Table \ref{zxs1x}
degenerates to the mere quadruple choice
as offered by Table \ref{bbs1x}.
The compatibility between the notation
used in the respective diagrams
(\ref{gewd2hu})
and (\ref{3humdi})
can be achieved via the above-mentioned identifications
like
${\cal J}_{0}\equiv  {\cal R}_0^{(1)}$,
${\cal K}_{0}\equiv  {\cal R}_0^{(2)}$,
${\cal H}_{textbook}\equiv  {\cal R}_0^{(3)}$ and
${\cal H}_{phys}\equiv {\cal R}_0\equiv {\cal R}_0^{(0)}$.
Next,
we believe that
for the present methodical purposes,
some persuasive results of the
comparison of
the separate strategies
can already be provided by the
study of the real toy-model matrices
of dimension three.
After all, even such ``first nontrivial'' matrices
could be interpreted as already forming
fairly reasonable and realistic
three-state quantum models.

The comparisons may then
start from the Dyson-map-factorization
formula (\ref{cocody}). Thus, let us
make the following one-parametric
illustrative choice of
 \be
 \Omega_3=\left[ \begin {array}{ccc}
  1&0&0\\\noalign{\medskip}r&1&r
  \\\noalign{\medskip}0&0&1\end {array} \right]
 \,,\ \ \ \ \ \
 \Omega_2=\left[ \begin {array}{ccc} 1&0&0
 \\\noalign{\medskip}s&1&0
 \\\noalign{\medskip}0&s&1\end {array} \right]
 \,,\ \ \ \ \ \
 \Omega_1=\left[ \begin {array}{ccc} 1&t&0
 \\\noalign{\medskip}0&1&t
 \\\noalign{\medskip}0&0&1\end {array} \right]\,.
 \label{samojed}
 \ee
This is a schematic ansatz which already yields
a more complicated but still just tridiagonal
three-parametric Dyson map product (\ref{cocody}),
 \be
 \Omega=\Omega_{321}=\left[ \begin {array}{ccc} 1&t&0
 \\\noalign{\medskip}r+s& \left( r+s \right) t+1+sr& \left( 1+sr \right) t+r
 \\\noalign{\medskip}0&s&ts+1
\end {array} \right]
 \ee
as well as the two two-parametric auxiliary Dyson sub-maps
used in diagram (\ref{3humdi}),
 \be
 \Omega_{32}=\left[ \begin {array}{ccc} 1&0&0
 \\\noalign{\medskip}r+s&1+sr&r
 \\\noalign{\medskip}0&s&1\end {array} \right]\,,
 \ \ \ \
 \Omega_{21}=
 \left[ \begin {array}{ccc} 1-ts&t&0
 \\\noalign{\medskip}0&1&t
 \\\noalign{\medskip}t{s}^{3}&-t{s}^{2}&ts+1\end {array} \right]\,.
 \ee
Obviously, using the knowledge
of the Dyson-map factors
(\ref{samojed})
we may immediately evaluate the
inner-product metrics entering the last line
of Table~\ref{bbs1x},
 \be
 Z_3=\left[ \begin {array}{lll} 1+{r}^{2}&r&{r}^{2}
 \\\noalign{\medskip}r&1&r
 \\\noalign{\medskip}{r}^{2}&r&1+{r}^{2}\end {array} \right]\,,\ \ \
 Z_3Z_2=\left[ \begin {array}{lll}
 1+({r}+{s})^{2}&(r+s)(1+sr)& \left( r+s \right) r
 \\\noalign{\medskip}(r+s)(1+sr)&(1+sr)^2+{s}^{2}&r+s{r}^{2}+s
 \\\noalign{\medskip}\left( r+s \right) r
 &r+s{r}^{2}+s&1+{r}^{2}\end {array} \right]
 \ee
and
 \be
 Z_3Z_2Z_1=
 \left[ \begin {array}{lll}
 1+({r}+{s})^{2}&\ldots&\ldots
 \\
 \noalign{\medskip}t\,(1+({r}+{s})^{2})
 +(r+s)(1+sr)&\ldots&\ldots
 \\
 \noalign{\medskip}\left( r+s \right)  \left( t+rts+r \right)
&\ldots&\ldots
\end {array} \right]\,.
 \ee
In the last item we only displayed
the first column of the matrix
because the other matrix elements would already
require more than one line.

We are prepared to finalize the comparison.
Our
message (that
the preferences may be expected to depend on the
structure of $H$)
may be accompanied by the
reference to paper \cite{BG} (showing that
the first-column BGI approach appeared best suited
for the very specific Buslaev's and Grecchi's
non-Hermitian but analytic model (\ref{ngo})
of Appendix B)
and also by the
reference to papers \cite{Dyson,Geyer}
where the authors advocated
the last-column DI option which proved optimal
for an efficient
numerical tractability
of a number of many-body quantum systems.

An important supplementary 
simplification of our argumentation will be based
on the idea that for any finite-dimensional matrix model
the concept of the complexity of an operator depends on the
choice of the basis. Really,
one can always find such an (in general, biorthonormal)
basis in which $H$ (or $\mathfrak{h}_1$, etc) is diagonalized.
In this sense, we decided to make our present, model-based
demonstration of the existence of the ``differences in the
complexity'' between the separate physical versions of the
eligible Hamiltonians in Table \ref{bbs1x} more transparent
by using the particular basis in which one postulates
the diagonality of the leftmost matrix $\mathfrak{h}_{321}$.

In the latter case, indeed, the correct and
physical inner-product metric is most elementary and, in fact,
trivial, $\Theta(\mathfrak{h}_{321})=I$.
This
enables us to
simplify our comparison of complexities by
starting
from a Hermitian matrix
$\mathfrak{h}_{321}$ chosen in its
simplest, diagonal and zero-parametric form.
Once we choose it in the following special version possessing
equidistant spectrum,
 \be
 \mathfrak{h}_{321}=\left[ \begin {array}{ccc} 1&0&0
 \\\noalign{\medskip}0&3&0
 \\\noalign{\medskip}0&0&5\end {array} \right]
 \ee
we may assign it, in the BGI scenario,
a phenomenologically
particularly appealing
truncated-harmonic-oscillator
interpretation.

In the opposite, maximal-shortcut DI-based
representation our choice
of the example then leads to the maximally
complicated
three-parametric
result of the Dyson map (\ref{preco}),
 \ben
 H= \left[ \begin {array}{lll}
 -2\,r{t}^{2}s-2\,tr-2\,{t}^{2}{s}^{2}-2\,t
 s+1&
 -2\,r{t}^{3}s-2\,{t}^{2}r-2\,{t}^{3}{s}^{2}-2\,t+2\,{t}^{2}{s}^{2}
 r+2\,rts&
 \ldots
 \\
 \noalign{\medskip}2\,rts+2\,r+2\,t{s}^{2}+2\,s&2\,r{t}^{2}s+2\,tr+2
  \,{t}^{2}{s}^{2}-2\,{s}^{2}rt+3-2\,sr&
  \ldots
 \\
 \noalign{\medskip}-2\, \left( r+s \right) s&
 -2\,rts-
 2\,t{s}^{2}+2\,s+2\,{s}^{2}r&
  \ldots
 \end {array}
 \right]\,.
 \een
%
This matrix (in which we omitted
the last column to fit the page) is,
in some sense, ``maximally'' non-Hermitian.
Any other isospectral and physical quasi-Hermitian
Hamiltonian
of Table \ref{bbs1x}
may be expected to be more user-friendly.

The latter expectations are manifestly confirmed
by the lengthy but straightforward linear-algebraic calculations
yielding the two-parametric isospectral partner Hamiltonian
 \be
 \mathfrak{h}_1=\mathfrak{h}_1(r,s)=\left[ \begin {array}{ccc} 1&0&0
 \\\noalign{\medskip}2\,r+2\,s&3-2\,sr&-2\,r
 \\\noalign{\medskip}-2\, \left( r+s \right) s&2\,s+2\,{s}^{2}r&2
\,sr+5\end {array} \right]
 \ee
(which already contains two vanishing matrix elements)
and, last but not least, the one-parametric
isospectral partner Hamiltonian
 \be
 \mathfrak{h}_{21}=\mathfrak{h}_{21}(r)=\left[ \begin {array}{ccc} 1&0&0
 \\\noalign{\medskip}2\,r&3&-2\,r
 \\\noalign{\medskip}0&0&5\end {array} \right]
 \ee
with as many as four vanishing matrix elements.

\begin{table}[h]

\caption{A sample test of the
benefits of alternative Hermitizations.
}
 \label{ccs1x} \vspace{.4cm}
\centering
\begin{tabular}{||c||c|c|c|c||}
    \hline \hline
   $N=3$&\multicolumn{1}{|c|}{BGI} &
    \multicolumn{3}{c||}{three shortcuts}\\
    \hline \hline
  Hamiltonian&
   $\mathfrak{h}_{321}$&$\mathfrak{h}_{21}$&$\mathfrak{h}_1$&$H$\\
   \hline
  ${\cal N}_{par}$&zero&one&two&three\\
${\cal N}_{zero}$ &6&4&2&0\\
\hline \hline
\end{tabular}
\end{table}

The results of
all of these observations are collected in Table \ref{ccs1x}. The
growth of the
complexity of the model is
mimicked there by the growth of the number ${\cal N}_{par}$
of its independently variable parameters.
The
desirable sparsity of the Hamiltonian matrix is
measured by the
number
${\cal N}_{zero}$ of its
vanishing
matrix elements.
We may conclude that
in spite of a highly schematic
nature of our illustrative example,
even the use of rather elementary criteria
indicates
the
marked differences between the alternatives.
Moreover,
both of our criteria
lead to the same
ordering of the recommended preferences,
with
an optimal Hermitization
being found in the
BGI approach.
In other words,
the test of benefits
is consistent in having both the physical Hamiltonians
and the corresponding inner-product metrics ordered,
in our methodically simplest illustration, from
the trivial BGI scenario up to the
opposite, least-friendly DI extreme.

For the sake of brevity we decided to
describe, in detail, just
the scenario in which
our toy model
appeared most complicated
in the DI approach of the last column of
Table~\ref{bbs1x}.
In a sharp contrast the model is trivial 
in the other, BGI
extreme of the
trivial-metric
BGI Hermitization as listed in the first column of
the same
Table.

In a way,
our present model
can be then perceived as a
simpler parallel of the
rather exceptional Buslaev's and Grecchi's
quantum system in which the BGI choice also appeared to be
optimal.
In our present matrix model,
nevertheless,
it would be possible to rotate the Hilbert-space basis
in a way reaching and simulating the preference of any other
Hermitization scenario.
Naturally,
in the less elementary unitary
quantum systems with non-Hermitian Hamiltonians,
the determination of a computationally optimal
Hermitization path will be less easy, in a way
strongly dependent on the properties of the
dynamical information as carried by the input
operator
$H$.

Obviously, we can summarize that 
our
unitary but still just hiddenly Hermitian
three-level quantum system is, for our present purposes,
sufficiently realistic.
In parallel, in the methodical context,
the model has been found sufficient to illustrate
the possibility of the sharp contrasts between
the extent of the
user-friendliness
(or, if you wish, of the
calculation complexities)
of the separate
Hermitization strategies.
The model
samples and reflects,
sufficiently distinctly, not only the qualitative features
of the DI and BGI extremes (with the latter one being
deliberately
constructed as the most user-friendly case)
but also the sufficiently persuasive differences
between the intermediate items in the Hermitization menu.
This seems to indicate that in applications it
truly might make sense to make use of the whole,
exhaustive list of all of
the partial-Dyson-map-mediated
Hermitizations of the form as
summarized
in Table \ref{zxs1x}.


\section{Discussion}

From the perspective of 
practical applicability of the present
reformulation of quantum theory
our motivation had two sources.
The first one may be found in
the multiple existing
difficulties of our current
understanding of the
quantization of gravity
\cite{Thiemann} and in their potential
tractability
by the mathematical
tools of 
QGQM \cite{ali}.
The second, related one 
lied in the study of the 
concept of the so called
minimal
length.
For example, the authors of
paper \cite{[12]}
claimed that
in such a context
the role
of the quasi-Hermiticity of the operators
of observables might be decisive.

For illustration they
decided to
treat the position operator
$X$ and the momentum operator $P$ as the
dynamical variables
which
are non-Hermitian but Hermitizable.
Subsequently, having used, in our terminology, the DI-type
version of the Hermitization,
their basic conjecture
(concerning the possibility of a {\em simultaneous\,} Hermitization
of {\em both\,} of the candidates $X$ and $P$ for an observable) 
appeared to be unfounded.
The main conclusion
of
our subsequent critical comments in \cite{arabky} or \cite{Lotor}
was that a guarantee of
a consistent
Hermitization of {\em several\,} non-Hermitian candidates $\Lambda_j$
for an observable
has to be understood as a
truly difficult open theoretical problem.

What followed was
a very recent partial resolution of the problem in
\cite{EPL}.
In that paper our message and construction were
based on
the $N-$term factorization of the
physical inner product metric $\Theta$.
Our main result was a constructive recipe by which
a multiplet of the observables
$\Lambda_j$ appeared obtainable in terms of the
metric factors $Z_n$.
Such an innovative DI-type result
offered in fact
an additional,
physics-based encouragement of our
present study.

One of the related and
currently open questions is
the search for a BGI-type complement
of the latter DI-type construction of the observables.
This is, obviously, directly connected with the 
applicability of
our present universal diagram (\ref{gewd2hu}).
One only has to keep in mind that
even the BGI extreme option
did not find applications 
in too many realistic models. 
In this sense the results obtained for
our toy model of section
\ref{sestasek} are encouraging.
Indicating that
the new horizons seem open by the
exhaustive classification
of (quasi-)Hermitizations via Eqs.~(\ref{smula})
and~(\ref{smulab}).
As long as the gap
between the DI and BGI
methodical extremes is now fully covered,
one might now return to the older
BGI-type results (cf., e.g., \cite{Mateo,Tenney})
and 
re-analyze the possibility of their
further simplification
in terms of
the new shortcuts of the BGI Hermitization.

The specification of an optimal choice 
of one of the present
upgrades of the general QHQM theory
is in fact a difficult problem.
Even the merits of 
our illustrative three-state model
of section~\ref{sestasek}
would quickly be lost after transition to larger
finite-dimensional matrices.
It is not clear in which way the necessary simplification
of the more realistic matrix models should proceed.
The general
criteria of the preference of a particular Hermitization
pattern are not yet known at present. 
In the future, the enumeration of the
merits of the choice
will certainly require a deeper
insight in the favorable 
properties of $H$. 

New sources of
difficulties will also emerge for the
infinite-dimensional Hilbert spaces
in which, for a given $H$, 
the inner-product
metric $\Theta=\Theta(H)$ need not exist
at all \cite{Dieudonne,Geyer,Lotor,Siegl}.
New ways towards the resolution of the
related mathematical problems
might start from the observation that
in the light of Eq.~(\ref{starstar})
the symbol $\Theta$
is just an abbreviation for a product of the
better-defined
Dyson-map operators. Besides their deeper
physical meaning (cf. Eq.~(\ref{repreco})), even
the mathematical conditions of their existence
may be perceivably less restrictive
(see Ref.~\cite{BG} for illustration once more).

From a purely pragmatic perspective
people should still search for the criteria of optimality.
One might feel encouraged by the fact that
even in the BGI extreme
based on the direct use and construction of the
fully factorized Dyson map (\ref{cocody})
the area of applicability is non-empty.
One may
return, once more, to Table \ref{os2x}
summarizing such a
trivial-inner-product strategy.
Now, the same Table can be re-read as admitting
the choice of
any $N>1$, with $k=N$
in Lemma \ref{lemmaone} and in Eq.~(\ref{smulab}).

\begin{table}[h]

\caption{Intermediate
Hermitizations
${\cal R}_{N}^{(0)}\to{\cal R}_{0}^{(i)}$
with  $i=1,2,\ldots,N-1$
.}
 \label{osNx} \vspace{.4cm}
\centering
\begin{tabular}{||c|c|c|c|c||}
    \hline \hline
    input:
& aim:& specification: & needed:
&calculated: \\ Hamiltonian
 & unitarity & path in diagram (\ref{gewd2hu})&  product in ${\cal R}_{0}^{(i)}$
& Hamiltonian in ${\cal R}_{0}^{(i)}$
 \\ $H \neq H^\dagger$ in ${\cal R}_{N}^{(0)}$
 & (in ${\cal R}_{0}^{(i)}$) & (cf. Eqs.~(\ref{smula}), (\ref{smulab}))
&(cf., e.g., Eq.~(\ref{3humdi}))
 &(cf., e.g., Table \ref{bbs1x})
 \\
\hline \hline
\end{tabular}
\end{table}

In this light
the main innovation and result of our present paper
may be seen
in the completion of systematics.
We described some basic features
of
a broad family of the new, ``intermediate''
Hermitizations
(summarized in Table \ref{osNx})
guaranteeing, equally well, the unitarity of the evolution
in the overall QHQM context.
On this background the message delivered by our
present paper is threefold. Firstly,
we reconfirmed the usefulness as well as a
not quite expected
methodical productivity
of the general
$N-$term factorization
of the
correct
physical Hilbert-space metric $\Theta$
(note that at $N=2$ one reobtains, along this line, even
the by far most popular
parity times charge ansatz for the metric).

Secondly,
via a deeper analysis of the closely related
composite-Dyson-map algebraic structures
we managed to offer a deeper interpretation of the
Buslaev's and Grecchi's ``exceptional''
anharmonic-oscillator example.
In a way
concerning the feasibility and usefulness of the 
similar systematic
non-DI Hermitization constructions, and in a way
opposing the existing BGI-related scepticism
we also proposed and described another, exactly solvable 
finite-matrix model 
for which the simplest Hermitization 
appeared to be provided strictly in its
BGI extreme.

Thirdly, last but not least,
we promoted the idea that
the factorization ansatzs may make the general
QHQM formulation of the theory
more user-friendly
and efficient
even between the two DI and BGI 
methodical extremes. The availability of 
the widened spectrum of Hermitizations
will enable one to
combine an optimal
partial amendment
of the physical inner product with a suitable complementary
partial
amendment
of the Hamiltonian and/or of the other relevant observables
in practice.

Concluding the discussion we believe that
the present completion of the list of optional Hermitizations
made the QHQM form of unitary quantum theory,
via an explicit and exhaustive enumeration of all
of the
alternative Hermitization patterns at a preselected fixed $N$,
more flexible and open not only to its
further methodical analyses
but also to its various ambitious phenomenological
applications -- for the freshmost illustration
see the discussion
of the potential relevance
of the QHQM language in quantum cosmology in
\cite{WDW}.

\section{Summary\label{disum}}

In our present paper
we felt challenged,
in the context of
the use of non-Hermitian Hamiltonians $H \neq H^\dagger$
in quantum mechanics of unitary systems,
by the dominance and prevailing preferences of the
most conventional
$N=1$
DI-Hermitization philosophy (see its compact summary
in Table~\ref{os1x}). We also
felt encouraged by
our preceding preliminary results concerning the possible
$N>1$ metric-factorization options: In
\cite{EPJP} we
described the most straightforward
amendment of the special, Hamiltonian-preserving
DI quasi-Hermitization
based on an $N-$term factorization of the
physical Hilbert-space
inner-product metric $\Theta$.

In our present continuation
of the analysis we were 
inspired by the Buslaev's and Grecchi's paper \cite{BG}.
We recalled their anharmonic-oscillator
Hamiltonian $H=H^{(BG)}$ as a methodical guide.
We emphasized that the
similar models are
characterized by the $N-$term
factorization (\ref{cocody})
of the underlyi8ng global Dyson map $\Omega$.
We imagined that the
existence of such a factorization
may
admit not only the brute-force
constructive BGI proof of the isospectrality
between
$H^{(BG)}$ and its
Hermitian avatar $\mathfrak{h}^{(BG)}$
but also a new family of
alternative Hermitizations.
In a constructive
and model-independent
manner we proposed and described
an exhaustive classification of
all of such $k-$step
Hermitizations with $0 \leq  k \leq N\,$ 
in Lemma \ref{lemmaone}.

In the language of diagrams
we re-classified the $(N+1)-$plet
of Hilbert spaces
${\cal R}_j^{(0)}$ which appears
in diagram (\ref{uzwedli})
as the mere upper-line
component of the more adequate,
much richer
and, in some sense, exhaustive
triangular classification diagram (\ref{gewd2hu}).
In the
diagram
the single-path DI Hermitization (\ref{usmultpl})
only emerges as
an extreme, $k=0$
special case.
The
gap in the methodical considerations of
paper \cite{EPJP}
is filled.
New perspectives are opened
in the analysis of models using
non-Hermitian observables. A number of problems
may now find new solutions mediated by the conjugate $N-$term
factorizations of the physical inner-product metrics
and of the related
Dyson maps.

\newpage

\section*{Appendix A. Hidden forms of Hermiticity\label{hifo}}

The numerical diagonalization of many realistic
self-adjoint Hamiltonians
$\mathfrak{h}=\mathfrak{h}^\dagger$
can be accelerated by their judicious
isospectral non-unitary preconditioning $\mathfrak{h} \to H$
(see Eq.~(\ref{preco})).
More than 60 years ago this has been revealed
by Freeman Dyson  \cite{Dyson}.
One only has to keep in mind that the manifest non-Hermiticity
$H \neq H^\dagger$ of $H$
is inessential because the
Hermiticity of
$ \mathfrak{h}= \mathfrak{h}^\dagger$
can be perceived as equivalent to the
quasi-Hermiticity of $H$ \cite{Dieudonne}.
In 1992, the arrow in (\ref{preco}) has been inverted
by Scholtz et al \cite{Geyer} (cf. Eq.~(\ref{repreco})).
A given, manifestly non-Hermitian $H$ with real spectrum
has been declared,
under certain not too restrictive technical conditions,
an acceptable Hamiltonian tractable as an
alternative representation of
$\mathfrak{h}$.

Even though the evolution of the wave functions then appeared
controlled by a manifestly non-Hermitian generator $H$ (cf. Eq.~(\ref{ostasta})),
the standard probabilistic interpretation of the theory
remained, due to the Dyson-map equivalence $\Omega$
and due to the Hermiticity of the partner $\mathfrak{h}$,
unchanged.
At present, the original
quasi-Hermitian formulation of quantum
mechanics (QHQM,  \cite{Geyer})
based on relation (\ref{starstar})
is considered, nevertheless, difficult to implement
(see, e.g.,
the arguments summarized on p. 1216 of newer review \cite{ali}).
For this reason
the ``Dyson-inspired'' (DI) Hermitization (\ref{repreco})
has been modified and made, more than twenty years ago,
technically more user-friendly.
A decisive simplification of the formalism has been achieved
within a slightly narrower framework
of the ${\cal PCT}-$symmetric quantum mechanics
(see its standard review \cite{Carl}).

One of the characteristic features
of the latter approach
is an
{\it ad hoc\,} two-term factorization
of
the metric: see our
comment on this feature in \cite{EPL}.
Still, even after the simplification,
every user of a
non-Hermitian but Hermitizable Hamiltonian $H$
had to
ask
the questions about
the compatibility of the non-Hermitian
$H$
with Stone theorem \cite{Stone}.
At present, it is already widely accepted that
the Hermitizations can be constructed as proceeding
either via a ``BGI-path'' transformation of the Hamiltonian
or via the mere ``DI-path'' amendment of the inner product
in ${\cal H}_{math}$.

In both cases, operator
$\mathfrak{h}$ is
much less user-friendly than $H$.
This explains why
the vast majority of the successful QHQM calculations are performed
in ${\cal H}_{math}$.
In the former case, nevertheless,
the Hermitization seems more difficult,
tractable in a narrower class of the solvable
or at least sufficiently elementary
models \cite{BG,Mateo}.
Still, the picture reconstructed in the
textbook space ${\cal H}_{textbook}$ may appear useful, say,
in semi-classical considerations \cite{Batal}.

In the latter, more robust case of the DI Hermitization
the user-friendly operator $H$ is kept unmodified.
The information about the
Hermiticity of $\mathfrak{h}$ in ${\cal H}_{textbook}$ only has
to be transferred
to the
working space ${\cal H}_{math}$
which is, by definition, manifestly unphysical.
Hence,
the transfer
leads
to the quasi-Hermiticity
constraint (\ref{starstar}).
This property of $H$ keeps the operator
tractable,
under certain additional technical assumptions
\cite{Geyer}, as hiddenly Hermitian.
Moreover, condition (\ref{starstar})
can be interpreted as inducing
an {\it ad hoc\,} change of inner product,
 \be
 \br \psi_a|\psi_b\kt \ \to \  \br \psi_a|\Theta|\psi_b\kt\,.
 \label{[6b]}
 \ee
Such a change can be read as a replacement of
the initial, manifestly unphysical manipulation space
${\cal H}_{math}$
by
its standard physical
partner space of states
${\cal H}_{phys}$. In it,
the
amendment of the Hermitian conjugation renders the initial
Hamiltonian $H$ self-adjoint in ${\cal H}_{phys}$
\cite{Dieudonne,Geyer}.

\section*{Appendix B.
The Buslaev's and Grecchi's oscillator}

In the specific illustrative model of paper \cite{BG}
(using, incidentally, a not too small
$N=N^{(BG)}=7$ in (\ref{cocody}) and (\ref{reaN})),
all of the recurrent
isospectrality-guaranteeing transformations
$ H_{0j}^{(BG)} \to  H_{0j+1}^{(BG)}$
were entirely elementary, mediated by
a non-numerical Fourier transformation or even by the
mere change of variables.
In this sense, the results were exceptional: Their authors
paid attention just to a special non-Hermitian Hamiltonian
 \be
 H=H_{\eta}({\rm i}g,j)=-\frac{d^2}{dx^2}+\frac{1}{4}V^{(BG)}_{(j,g,\eta)}(x)
 \neq H^\dagger
 \,,\ \ \ \ \  x \in \mathbb{R}
 \,
 \label{ngo}
 \ee
containing an unusual, complex and asymptotically wrong-sign
anharmonic-oscillator potential
 \be
 V^{(BG)}_{(j,g,\eta)}(x)=
 \frac{j^2-1}{r^2_{\eta}(x)}+
 {r^2_{\eta}(x)}{}-
 {g^2\,r^4_{\eta}(x)}{}\,,
 \ \ \ \ \ r_{\eta}(x)=x-{\rm i}{\eta}\,,\ \ \ \ \eta>0\,,\ \ \ j>0
 \,
 \label{pogro}
 \ee
(cf. equation Nr. 16 in {\it loc. cit.}). These authors
identified the model as ${\cal PT}-$symmetric (parity-time-symmetric)
and proved that its spectrum is discrete, real and bounded from below.
The main idea of the proof lied in the observation
that
Hamiltonian
$H_{\eta}({\rm i}g,j)$
(which is non-Hermitian in the conventional Hilbert space
$L^2(\mathbb{R})\,\equiv \,{\cal H}_{math}$)
is equivalent (i.e., isospectral)
to another operator
 \be
 Q(g,j)=-\frac{d^2}{dx^2}
  +\mathfrak{v}^{(BG)}_{(j,g)}(x)=Q^\ddagger(g,j)
 \,,\ \ \ \ \  x \in \mathbb{R}
 \,
 \label{vchupe}
 \ee
with a real and
confining double-well potential
 \be
 \mathfrak{v}^{(BG)}_{(j,g)}(x)=(gx-1)^2\,x^2+j\,(1/2-gx)\,.
 \label{ovchupe}
 \ee
The double well model
was
safely self-adjoint in the
Hilbert space $L^2(\mathbb{R})\,\equiv \,{\cal H}_{textbook}$
and admitted, therefore, the standard
unitary-evolution quantum-mechanical interpretation
(cf. equation Nr. 17 and Theorem~5 in {\it loc. cit.}).

At its time, the result was treated as a
mere mathematical curiosity
\cite{Gprco}. The situation has only changed
when, a few years later,
Bender with Milton \cite{BM},
Bender with Boettcher \cite{BB}
and a few other teams of authors \cite{CJT,Rafa}
pointed out
that the similar non-Hermitian but
${\cal PT}-$symmetric quantum Hamiltonians $H\neq H^\dagger$
with real spectra
might be of an immediate theoretical as well as
phenomenological interest
(see also reviews \cite{ali,Carl,SIGMA,book}).

In this context
the birth of the motivation of our present considerations dates back
to the year 2000 when
Vincenzo Grecchi
\cite{Gprco} recalled
his collaboration
with Vladimir Buslaev and when he
pointed out that
at least some of the
topical open questions
concerning the ${\cal PT}-$symmetric quantum systems
have already been answered
in
\cite{BG}.
One of these answers appeared
in the proof of Theorem 5. In it,
the authors
used the factorization~(\ref{reaN}) and they
filled the gap between
the respective
initial and final operators $H^{}_{\eta}({\rm i}g,j)$ and $Q(g,j)$
by a set of auxiliary isospectral operators,
 \be
 H^{}_{\eta}({\rm i}g,j) \sim
 K^{4}_{\eta}({\rm i}g,j) \sim \ldots  \sim
 Q^{6}_{\eta}(g,j) \sim  Q(g,j)\,.
 \label{illu}
 \ee
These operators
formed a set
of equivalent candidates for the Hamiltonian defined
by equations Nr. 19 - 24 in \cite{BG}. They
were allowed to fail to be Hermitian or
${\cal PT}-$symmetric. Their
mutual
isospectrality
was comparatively easily deduced
from the underlying
recurrent
definitions of the operator-transformation form
$K^{4}_{\eta}({\rm i}g,j)={\cal S}_1
H^{}_{\eta}({\rm i}g,j){\cal S}^{-1}_1
$, etc.
All of the elements of
multiplet (\ref{illu})
happened to be just ordinary linear differential operators
defined, in closed form, analytically.


\section*{Appendix C. Factorized metrics}

In the three-space
diagram (\ref{rexhum}) the realization
of both of the BGI and DI Hermitizations needs just
$N=1$ in Eq.~(\ref{cocody}).
At the same time, the
DI-path strategy itself
is already known in the generalized form using
any $N$ \cite{EPJP,EPL}.
This is one of the reasons why the more flexible $N-$step
DI-path Hermitization might seem preferable in applications.
Indeed, it requires just the
solution of Eq.~(\ref{starstar}) for metric
$\Theta=\Theta(H)$
(see also a few related comments on p.~1216
in review \cite{ali}, and
an outline of
the construction methods
to be found, e.g.,
in section Nr. 4 of the same review).

\subsection*{C. 1. $N=2$}

A remarkable feature of the Hamiltonians $H$
which are quasi-Hermitian {\it alias\,}
``Hermitian in disguise'' is that
their popularity
started to grow only {\it after\,}
the
theory was complemented
by the innocent-looking requirement of ${\cal PT}-$symmetry
(see also the similar comment in Appendix~A).
This led to a decisive innovation of
the search for the Hermitizing
metric $\Theta$ which has been found facilitated
by its pre-factorization,
 \be
 \Theta=Z_2\,Z_1\,.
 \label{je2}
 \ee
The right
factor
was
kept Hamiltonian-dependent and treated, initially at least, as a
charge, $Z_1=Z_1(H)=Z_1^\dagger={\cal C}$.
The left one was initially
defined as the conventional parity,
$Z_2=Z_2^\dagger={\cal P}$.
Some authors criticized
ansatz (\ref{je2}) as
not too well motivated
(see, e.g.,
pp.~1198 and 1232 in review~\cite{ali}).
We oppose the criticism
emphasizing that
the factorization contributed to
the appeal of QHQM
significantly.

We admit that
the
motivation of
ansatz (\ref{je2}) seems
mysterious and that it deserves a clarification.
For the purpose we
reflected a part of the criticism and we
turned attention to a possible generalization of the ansatz.
In \cite{PLA}, for example, we reinterpreted
the charge
$Z_1$ 
as the mere independent
non-Hermitian but
${\cal PT}-$symmetric 
observable.
In another, model-independent study of the
problem in \cite{EPL} it has been reconfirmed
that the apparent mystery
of ansatz (\ref{je2}) may be given a very natural clarification
based on
a less specific choice of the
factors $Z_j$.
In this manner a better
balance has been
established between
the roles played by the factors
$Z_1$ and
$Z_2$.

The balance has been achieved
at the expense of introduction
of another, third, ``intermediate'' inner-product
space ${\cal H}_{int}$.
In terms of diagrams we just proposed
the upgrade (\ref{wd2humdi}) of diagram~(\ref{krexhu}).
The presence of the two factors in (\ref{je2})
and of the two
steps in diagram (\ref{wd2humdi})
implies that
the metric $Z_2$ need not necessarily be
positive definite.
In our present paper, nevertheless,
only the simpler Hilbert-space scenario characterized by
the positive definite factors $Z_1$ and $Z_2$
will be considered
for the sake of brevity (see also
the extensive discussion of such an 
option in \cite{shendr}).

The two-term
factorization of the metric (\ref{je2}) may be assigned
a deeper mathematical meaning.
In~\cite{PLA} we emphasized that
for consistency reasons it would be sufficient when
operator
$Z_1$ is merely required
quasi-Hermitian
with respect to another, auxiliary
``partial metric'' $Z_2$.
This means that
the related modified, positively definite
``anomalous observable charge'' $Z_1$ need not be Hermitian
in ${\cal H}_{math}$, $Z_1 \neq Z_1^\dagger$.
Due to
the Hermiticity
of the positive-definite operator $Z_2=Z_2^\dagger >0$
the most fundamental condition of the
Hermiticity
of the factorized
metric (\ref{je2})
may be now most easily satisfied because it
is equivalent to the
single condition
 \be
 Z_1^\dagger\,Z_2=Z_2\,Z_1\,
 \label{rerea}
 \ee
representing the weakened requirement of the
$Z_2-$mediated quasi-Hermiticity
of $Z_1$.
Thus, one can conclude that
it is possible to work with the Hermitization
diagram (\ref{wd2humdi}) which contains the
inner-product space
${\cal H}_{int}$ endowed with the new auxiliary and positive definite
metric $Z_2$.

In the intermediate Hilbert space every ket vector $|\psi\kt$
becomes associated
with an unusual form of the Hermitian-conjugate
bra vector $\br \psi|\,Z_2$. This means that
the information provided by
the first line of
diagram~(\ref{wd2humdi}) may be made
more explicit, including the
``local'', inner-product-dependent bra-vector
Hermitian conjugates,
 \be
 \label{u2hum}
  \ba
  \ \ \ \ \ \ \ \ \ \ \
   \begin{array}{|c|}
 \hline
 {\rm   } |\psi_{m}\kt,
  \ {\rm  }
 \br \psi_{n} | \\
 {\rm for }\ {\cal H}_{math}\\
 \hline
 \ea
 \
 \stackrel{\rm step\ 1}{ \longrightarrow }
 \
 \begin{array}{|c|}
 \hline
 {\rm  } |\psi_{m}\kt,
  \ {\rm  }
 \br \psi_{n} |\,Z_2 \\
 {\rm for }\ {\cal H}_{int}\\
 \hline
 \ea
 \
 \stackrel{\rm step\ 2 }{ \longrightarrow }
 \
 \begin{array}{|c|}
 \hline
 {\rm   } |\psi_{m}\kt,
  \ {\rm   }
 \br \psi_{n} |\,Z_2\,Z_1\\
 {\rm for }\ {\cal H}_{phys}\\
 \hline
 \ea\ .
\ea
 \ee
In such a two-step DI-path Hermitization scenario,
the description of quantum
dynamics is still based on the solution of
Schr\"{o}dinger equation using {\em the same\,}
Hamiltonian operator $H$.

The idea leads to a reformulation of quantum mechanics
in which the same Hamiltonian $H$ is assigned
the three
different,
inner-product-dependent Hermitian conjugate
forms. In~\cite{PLA}
they were denoted by the three
different dedicated symbols, viz., as
$H^\dagger$ in ${\cal H}_{math}$, as
$H^\ddagger$ in ${\cal H}_{int}$, and as
$H^\sharp$ in ${\cal H}_{phys}$.
Later,
all of the necessary formulae were
represented in ${\cal H}_{math}$ so that
the introduction of the
special dedicated superscript in $H^\ddagger$ or in $H^\sharp$
can be declared
redundant. For this reason,
the exclusive representation
of the system in ${\cal H}_{math}$ is
also used in our present paper.


\subsection*{C. 2. $N=3$}

In an immediate
generalization
of the above-outlined two-factor scheme
let us consider
the three-term ansatz
 \be
 \Theta=Z_3\,Z_2\,Z_1\,.
 \label{je3}
 \ee
Its introduction
merely
leads to the replacement
of the four-space Hermitization pattern (\ref{wd2humdi})
by its five-space analogue (\ref{zwd3humdi}).
On the technical level
the Hermiticity property of parity is then transferred to another
operator,
$Z_3=Z_3^\dagger$.
The
quasi-Hermiticity
(\ref{rerea}) has to be upgraded yielding the doublet of
the necessary and sufficient constraints,
 \be
  Z_2^\dagger\,Z_3=Z_3\,Z_2\,, \ \ \
  Z_1^\dagger\,(Z_3\,Z_2)=(Z_3\,Z_2)\,Z_1
    \,.
 \label{3rerea}
 \ee
In practice, therefore,
the ``dynamical input'' knowledge
of the positive definite operator $Z_3$ and of
any positive definite Hermitian operator
$Y_3=Z_3\,Z_2$ enables us to reconstruct $Z_2$
in ${\cal H}_{math}$
while, similarly, another piece of
the ``dynamical input'' knowledge
of a positive definite Hermitian operator
$X_3=Z_3\,Z_2\,Z_1$ enables us to reconstruct $Z_1$.
Ultimately, we may deduce and prove the observability
status of
the operators $\Lambda_1=Z_1$ and  $\Lambda_2=Z_2\,Z_1$ \cite{EPL}.
The DI-path sequence of
Eq.~(\ref{u2hum}) can be replaced by the three-step
Hermitization pattern,
\be
 \label{ff3humdi}
  \ba
     \begin{array}{|c|}
 \hline
 {\rm   } |\psi_{m}\kt,
 \br \psi_{n} | \\
  {\rm for\ }{\cal R}_{3}^{(0)}\\
 \hline
 \ea
 \
 \stackrel{{\rm step\ 1}}{ \longrightarrow }
 \
 \begin{array}{|c|}
 \hline
 {\rm  } |\psi_{m}\kt,
 \br \psi_{n} |\, Z_3 \\
 {\rm for\ }{\cal R}_{2}^{(0)}\\
 \hline
 \ea
 \
 \stackrel{{\rm  step\ 2}}{ \longrightarrow }
 \
 \begin{array}{|c|}
 \hline
 {\rm   } |\psi_{m}\kt,
 \br \psi_{n} |\, Z_{3}\,Z_2 \\
 {\rm for\ }{\cal R}_{1}^{(0)}\\
 \hline
 \ea
 \
 \stackrel{{\rm  step\ 3}}{ \longrightarrow }
 \
 \begin{array}{|c|}
 \hline
 {\rm   } |\psi_{m}\kt,
 \br \psi_{n} |\, Z_3\,Z_2\,Z_1 \\
 {\rm for\ }{\cal R}_{0}^{(0)}\\
 \hline
 \ea\,.
  \ea
  \ee
Both of the two- and three-step DI-path
Hermitizations remain conceptually
the same.
The transition to the general $N-$step DI-path
Hermitization is immediate
and may be found described, thoroughly, in \cite{EPJP}.

\end{document}